\documentclass[twocolumn,usenatbib]{mnras}

\usepackage{graphicx}
\usepackage{amsmath}

\usepackage{xcolor}

\def\pmb#1{\setbox0=\hbox{#1}%
    \kern-.025em\copy0\kern-\wd0
    \kern.05em\copy0\kern-\wd0
    \kern-.025em\raise.0433em\box0}

\def\ltsima{$\; \buildrel < \over \sim \;$}
\def\gtsima{$\; \buildrel > \over \sim \;$}
\def\simlt{\lower.5ex\hbox{\ltsima}}
\def\simgt{\lower.5ex\hbox{\gtsima}}
\def\p2Y{\;_2Y}
\def\m2Y{\;_{-2}Y}

\def\mk2{\mu {\rm K}^2}
\def\Planck{\it Planck\rm}
\def\Plancks{\it Planck \rm}
\def\LCDM{$\Lambda$CDM}

\newcommand{\camspec}{{\tt CamSpec}}

\newcommand{\flamingo}{{\tt FLAMINGO}}

\def\pmb#1{\setbox0=\hbox{#1}%
     \kern-.025em\copy0\kern-\wd0
     \kern.05em\copy0\kern-\wd0
     \kern-.025em\raise.0433em\box0}

\begin{document}

\title[Thermal Sunyaev-Zeldovich Effect]{The Power Spectrum of the
  Thermal Sunyaev-Zeldovich Effect}

\author[George Efstathiou and Fiona McCarthy]{George Efstathiou$^{1,2}$ and Fiona McCarthy$^{3,1,4}$\\
[1] Kavli Institute for Cosmology Cambridge, Madingley Road, Cambridge, CB3 OHA, UK. \\
[2] Institute of Astronomy, Madingley Road, Cambridge, CB3 OHA, UK.\\
[3] DAMTP, Centre for Mathematical Sciences, University of Cambridge, Wilberforce Road, Cambridge CB3 OWA, UK.\\
[4] Center for Computational Astrophysics, Flatiron Institute, 162 5th Avenue, New York, NY 10010 USA.
}

\maketitle

\begin{abstract}
  The power spectrum  of unresolved  thermal Sunyaev-Zeldovich (tSZ)  clusters is extremely sensitive to the amplitude of the matter fluctuations. This paper present an analysis of the tSZ power spectrum using temperature power spectra of the cosmic microwave background (CMB)  rather than  maps of the Compton y-parameter.  Our analysis is  robust and insensitive to 
  the cosmic infrared background. Using data from \Planck, 
  and higher resolution CMB data from the Atacama Cosmology Telescope
  and the South Pole Telescope, we find strong  evidence that the tSZ spectrum has a shallower slope and a  much lower amplitude at multipoles $\ell \simgt 2000$ compared to the predictions of the baseline \flamingo\ hydrodynamic simulations of the \LCDM\ cosmology.
  Recent results on CMB lensing, cross-correlations of CMB lensing with galaxy surveys and full shape analysis of 
  galaxies and quasars from the Dark Energy Spectroscopic Instrument  suggests that this discrepancy cannot be resolved by lowering the amplitude of the matter fluctuations. An alternative possibility is that the  impact of baryonic feedback in the
  baseline \flamingo\ simulations is underestimated.
  
\end{abstract}


\section{Introduction}
\label{sec:Introduction}

The thermal Sunyaev-Zeldovich (tSZ) is caused by the inverse Compton scattering of cosmic microwave background  photons with the electrons in the hot atmospheres of groups and clusters of galaxies \citep{SZ:1972}. The tSZ effect can be disentangled from the primordial blackbody CMB anisotropies via its distinctive spectral signature, offering a potentially powerful probe of structure formation. Furthermore,  it has long been known that the integrated tSZ signal from clusters depends sensitively on the amplitude of the matter fluctuation spectrum \citep{Cole:1988, Komatsu:1999, Komatsu:2002}.

Let us define the Compton y-parameter  seen on the sky in direction $\pmb{$\hat  l$}$ by the line-of-sight integral 
\begin{equation}
   y = \int n_e  {kT_e \over m_e c^2} \sigma_T dl,  \label{equ:tSZ1}
\end{equation}
where $n_e$ and $T_e$ are the electron density and temperature and $\sigma_T$ is the Thomson cross-section. At frequency $\nu$, the tSZ effect produces a  change in the thermodynamic temperature of the CMB of 
\begin{subequations}
\begin{equation} 
   {\Delta T \over T_{\rm CMB} }  = f(x) y,   \label{equ:tSZ2}
\end{equation}
where\footnote{We use the notation $h_p$ for the Planck constant to distinguish it from the dimensionless Hubble parameter.}
\begin{equation}
 f(x) =  x {(e^x + 1) \over (e^x - 1)} - 4, \quad x\equiv {h_p\nu \over kT_{\rm CMB}}, \label{equ:tSZ3}
\end{equation}
\end{subequations}
\citep[see e.g.][for  review of the tSZ effect]{Carlstrom:2002}. 

\cite{Komatsu:2002} made reasonable assumptions concerning the pressure profiles of clusters (discussed in more detail below) and integrated over the cluster mass function to  make theoretical predictions for the tSZ power spectrum, $C^{yy_{\rm pred}}_\ell$,  expected in a \LCDM\  cosmology.  They found the following scaling with cosmological parameters:
\begin{equation}
C^{yy_{\rm pred}}_\ell \propto \sigma_8^{8.1} \Omega_m^{3.2} h^{-1.7}, \label{equ:tSZscaling1}    
\end{equation}
where  $\sigma_8$ is the root mean square linear  amplitude of the matter fluctuation spectrum in spheres of radius $8 h^{-1} {\rm Mpc}$ extrapolated to the present day, $\Omega_{\rm m}$ is the present day matter density in units of the critical density
and $h$ is the value of the Hubble constant $H_0$ in units of $100 \ {\rm km}{\rm s}^{-1}{\rm Mpc}^{-1}$. We can rewite Eq.~\ref{equ:tSZscaling1} as 
\begin{equation}
C^{yy_{\rm pred}}_\ell \propto (S_8 \omega_m^{-0.1})^{8.1},  \label{equ:tSZscaling2}    
\end{equation}
where $S_8$ is the parameter combination $S_8 = \sigma_8 (\Omega_m/0.3)^{0.5}$, which is accurately measured in cosmic shear surveys,  and $\omega_m = \Omega_m h^2$ measures the physical
density of matter in the Universe. The parameter $\omega_m$ is determined  very accurately from the acoustic peak structure of the CMB temperature and polarization power spectra in the minimal 6-parameter  \LCDM\ cosmology and is insensitive to 
simple extensions beyond \LCDM\  \citep[][hereafter P20]{Planck_params:2020}.   Thus the amplitude of the tSZ power spectrum is expected to depend sensitively on the  $S_8$ parameter.

Observations of the tSZ effect therefore have a bearing on  the discrepancy between the value of $S_8$ determined from the CMB and the values inferred from weak galaxy lensing surveys \citep[e.g.]{Hikage:2019, Asgari:2021, Amon:2022, Secco:2022} which has become known as the `$S_8$-tension'. The discrepancy is at the level of $\sim 1.5-3 \sigma$, depending on the specific
weak lensing survey,  choices of scale-cuts,  model for intrinsic alignments  and
assumptions concerning baryonic physics \citep{AmonEfstathiou:2022, Preston:2023, DES+KIDS:2023}. Although this  is not strongly significant, an $S_8$ tension has been reported since the  early days of weak lensing surveys \citep{Heymans:2012, MacCrann:2015}. The question of whether cosmic shear measurements require new physics beyond \LCDM\ is unresolved and remains a topic of ongoing research.

This paper is motivated by the analysis of the \flamingo\ suite of cosmological hydrodynamical simulations   presented in \cite{McCarthy:2023} (hereafter M23). (See \cite{Schaye:2023} for an overview of the \flamingo\ project and a description 
of the simulations used in M23.) The main aim of M23 is to assess the impact of baryonic feedback of various physical quantities sensitive to the $S_8$ parameter, including galaxy shear two-point statistics and the tSZ power spectrum. The `sub-grid' feedback 
prescriptions used in the baseline \flamingo\ simulations are constrained to match the present day galaxy stellar mass function and the gas fractions observed in groups
and clusters of galaxies.  One of their most striking results concerns the tSZ power spectrum. They argue that the tSZ power spectrum is dominated by massive clusters and is therefore insensitive to small variations of the baryonic feedback model around the baseline.
 Yet their simulation predictions based on a \Planck-like \LCDM\ cosmology have a much higher amplitude than the tSZ power spectrum inferred by \cite{Bolliet:2018} (hereafter B18) from the \Planck\ map of the tSZ effect \citep{Planck_Ymap:2016} and with the tSZ amplitude inferred at high multipoles from observations with the South Pole Telescope (SPT) 
\citep{Reichardt:2021}. Since
the amplitude of the tSZ signal is strongly dependent on the $S_8$ parameter (Equ. \ref{equ:tSZscaling2}) M23 conclude
that a new  physical mechanism is required to lower the value of $S_8$ below that of the \Planck\ \LCDM\ cosmology. \footnote{Note that this discrepancy was first highlighted by \cite{McCarthy:2014} who showed, using an early series of numerical hydrodynamic simulations,  that  the tSZ power spectrum predicted assuming  the  \Planck~cosmology had a higher amplitude  than
inferred from ACT and SPT \citep{Reichardt:2012,  Sievers:2013}.} This
is potentially an important result since it presents evidence for an $S_8$ tension independent of cosmic shear surveys
using a statistic that is claimed to be
insensitive to baryonic feedback processes.

We reassess the conclusions  of M23 in this paper. As  discussed  in Sect.~\ref{sec:motivation} and in more detail in Sect. \ref{sec:ymaps}, 
power spectra computed from \Planck-based Compton y-maps are strongly contaminated by several  components
which must be known and subtracted  to high accuracy to infer a tSZ power spectrum. Section \ref{sec:tSZspec} presents a much simpler power-spectrum based approach applied to \Planck\  data. Our analysis is  designed to isolate the
unresolved tSZ effect and the white noise contribution
from radio point sources  from the cosmic infrared background (CIB),  which is poorly known at frequencies $\simlt 217$ GHz. The amplitude of the radio source power spectrum can be constrained from deep number counts, breaking the degeneracy between the tSZ and radio source power spectra.
We also present results of power spectrum analyses at high multipoles using data from SPT and      the Atacama Cosmology Telescope (ACT) \citep{Sievers:2013, Das:2014, Choi:2020}. Finally,  we combine data from \Planck, ACT and SPT to reconstruct the shape of the tSZ power spectrum over the multipole range $\ell \sim 200 \mbox{--} 7000$.  Our conclusions are summarized in Sec.~\ref{sec:conclusions}.

\section{The motivation for this paper}
\label{sec:motivation}

\begin{figure}
	\centering
	\includegraphics[width=85mm, angle=0]{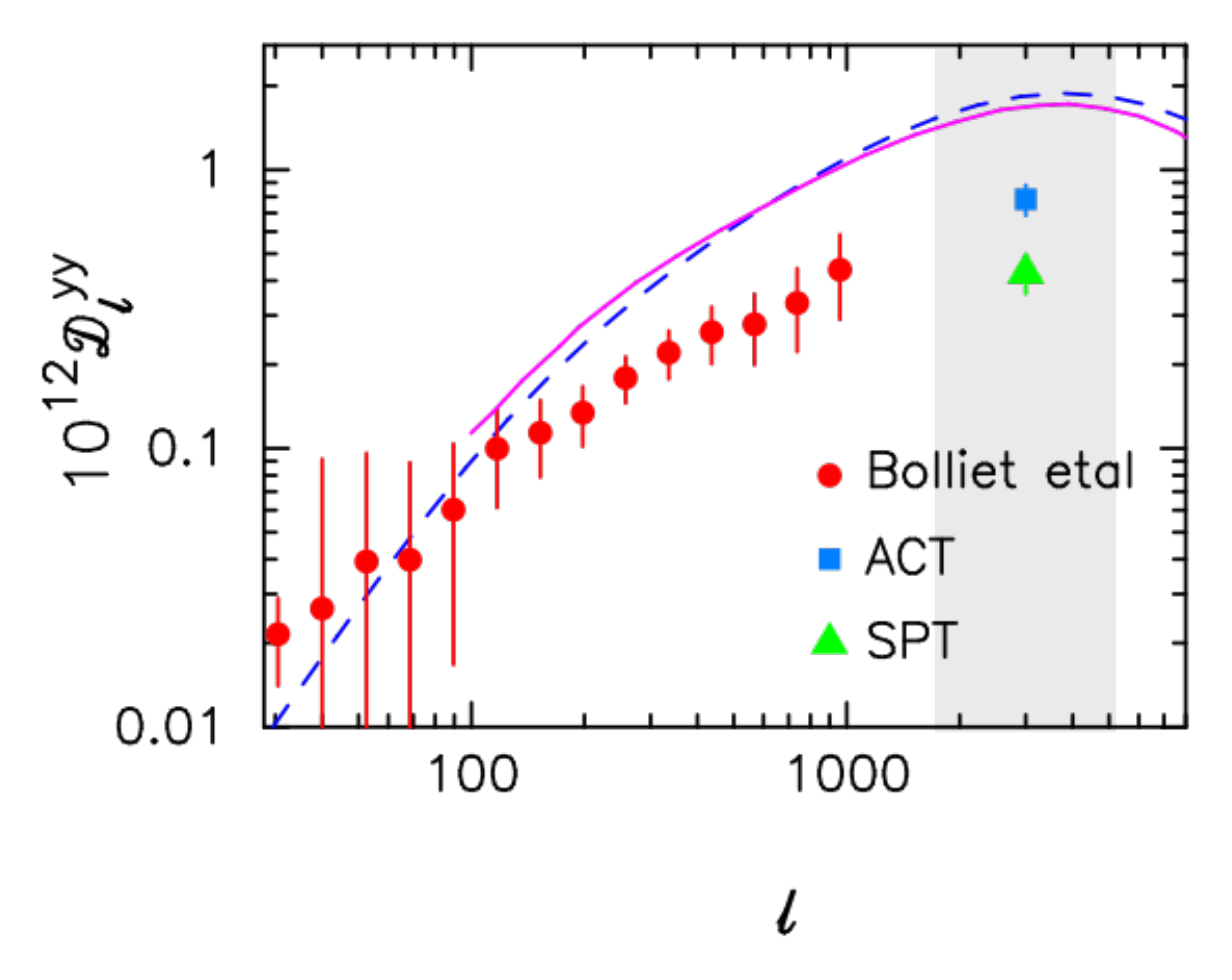} 
	\caption{The red points show estimates of the tSZ power spectrum from B18, together with $1\sigma$ errors.
           The blue and green points show the amplitudes of template tSZ power spectra at $\ell =3000$ inferred from high resolution ground based
          CMB power spectra measured by the ACT and SPT collaborations \citep{Choi:2020, Reichardt:2021}
          The grey band is included to highlight the fact that the  ACT and SPT measurements are model dependent amplitudes rather than measurements at $\ell = 3000$. The purple line shows the tSZ spectrum determined from a \flamingo\ simulation of the \Planck\ \LCDM\  cosmology assuming their default `sub-grid' parameters. The dashed blue line shows a simple one-halo model described in the text that is  designed to match the \flamingo\ results.    }

	\label{fig:flamingo}

\end{figure}

Figure \ref{fig:flamingo} is based on Fig. 5 from M23.  The red points show the tSZ power spectrum inferred by B18\footnote{
Specifically,  B18 use a cross-spectrum of the Needlet Internal Linear Combination \citep[NILC][]{Delabrouille:2009}  y-map constructed from the first half of the \Planck\ data and the Modified Internal Linear Combination Algorithm \citep[MILCA][]{Hurier:2013} y-map from the second half of the \Planck\ data.} from the 
\Planck\ all-sky maps of the Compton y-parameter (which we will refer to as y-maps) available from the Planck Legacy Archive\footnote{https://pla.esac.esa.int} (PLA) \citep{Planck_Ymap:2016}.

\begin{figure*}
	\centering
	\includegraphics[width=190mm, angle=0]{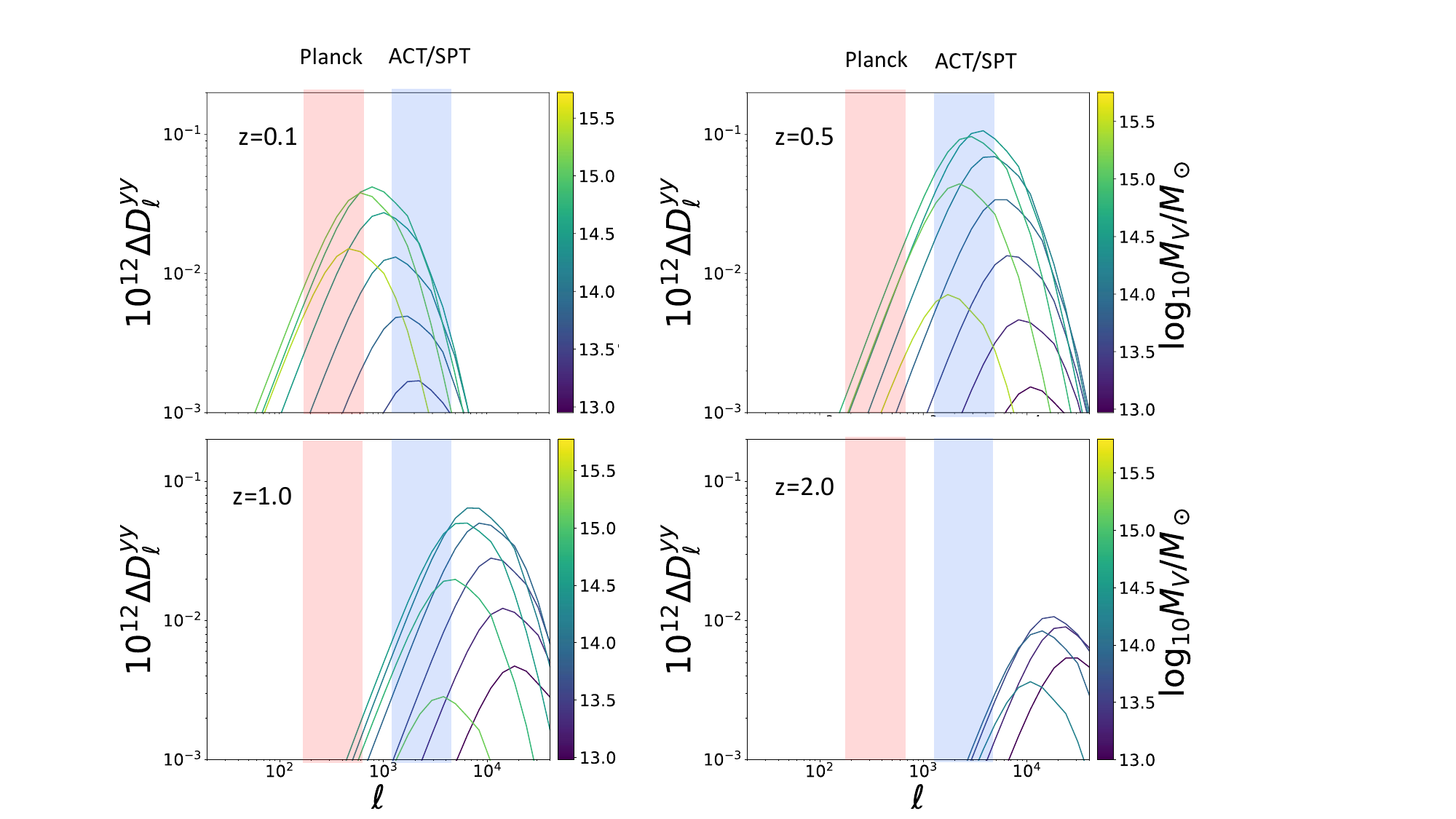} 
	\caption{The contribution to the tSZ power spectrum, computed from the one-halo model described in the text, plotted as a function of virial cluster mass $M_V$ (measured in $M_\odot$), redshift and multipole. The evolution parameter $\epsilon$ in Eq.~\ref{equ:P1} has been set to $\epsilon = 1$ in this example, leading to the sharp decline in power at $z\simgt 1$ (see Sect~\ref{sec:conclusions}).  The shaded regions show the approximate range of multipoles probed by \Planck,  ACT and SPT. 
	\label{fig:evolution}.}

\end{figure*}

The error bars show $1\sigma$ errors as reported in B18. The purple line shows the \flamingo\ results from M23 for a simulation of the \Planck\ \LCDM\ cosmology using  their default feedback
parameters  (solid green line  in the upper right hand plot in Fig. 5 from M23).  The two points at multipoles of $\ell \approx 3000$ show the amplitude of the tSZ power spectrum inferred from SPT \citep{Reichardt:2021} and from ACT \citep{Choi:2020}. These measurements at high multipoles are fundamentally different from the B18 analysis
since they are based on fits of a parametric foreground model to temperature power spectra   whereas  B18 
infer the tSZ power spectrum  from  y-maps. Although the high-multipole results  are usually plotted at $\ell = 3000$,  as in Fig.~{\ref{fig:flamingo}, it is important to emphasise that these points are not measurements of the tSZ
  signal at $\ell = 3000$. They give the amplitude of an assumed  tSZ template spectrum at $\ell=3000$. To emphasise this difference, we have superimposed a shaded area over these points
  to signify qualitatively that a range of angular scales contributes to  the ACT and SPT measurements. Note further that the ACT and SPT amplitudes appear to  differ by
  $\sim 2\sigma$. The consistency of the ACT and SPT tSZ results and the
  exact multipole ranges sampled by these experiments will be made more precise in Sect.~\ref{sec:tSZspec}.  

  One can see that the  \flamingo\ curve fails to match the B18 points by a wide margin. M23 also ran a set of simulations,  labelled LS8,  which have \Planck-\LCDM\ parameters except that the amplitude of the linear fluctuation spectrum was reduced to gve a low value of $S_8 = 0.766$ at the present day to match the weak lensing results reported by \cite{Amon:2023}. The LS8 cosmology clips the upper ends of the B18 error bars and so  provides a better  fit to the data than the \Planck\ \LCDM\ 
  cosmology. The LS8 model moves in the right direction but does not lower the tSZ amplitude enough to explain the B18 results.
  Furthermore, the LS8 model  fails to match the ACT and SPT point by many standard deviations.
  Evidently,  simply lowering the amplitude of the fluctuation spectrum to match weak lensing measurements
  cannot reconcile the simulations with the ACT and SPT data points shown in Fig.~\ref{fig:flamingo}.

The dotted blue line in Fig.~\ref{fig:flamingo} shows a one-halo model \citep{Komatsu:2002}
for the tSZ power spectrum computed as described in
\cite{Efstathiou:2012}   
for the \Planck-like \LCDM\ cosmological parameters adopted in M23. The electron
pressure profile was assumed to follow the `universal' pressure profile of
\cite{Arnaud:2010}
\begin{equation}
P_{\rm e}(x) = 1.88\left [{ M_{500} \over 10^{14}h^{-1} M_\odot} \right ]^{0.787} p(x) E(z)^{{8\over 3}-\epsilon}h^2 {\rm eV}
\ {\rm cm}^{-3}, \label{equ:P1}
\end{equation}
where 
\begin{equation}
p(x) ={ P_0 h^{-3/2}  \over (c_{500} x)^{\gamma} (1 + [c_{500}x]^{\alpha})^{(\beta - \gamma)/\alpha}},  \label{equ:P2}
\end{equation}
with the parameters $P_0 = 4.921$, $c_{500} = 1.177$, $\gamma = 0.3081$, $\alpha = 1.051$, $\beta=5.4005$ and $x=r/R_{500}$.   Here $R_{500}$ is the radius at which the cluster has a density contrast of 500 times the critical density at the redshift of the cluster, $M_{500}$ is the mass of the cluster within $R_{500}$, and the function $E(z)$ in (\ref{equ:P1}) is the ratio
of the Hubble parameter at redshift $z$ to its present value,
\begin{equation}
E(z) = \left[ (1 - \Omega_\Lambda) (1+z)^3 + \Omega_\Lambda \right]^{1/2}. \label{equ:P3}
\end{equation} 
The scaling $E(z)^{8/3}$  in Eq. (\ref{equ:P1}) assumes  self-similar evolution and the parameter $\epsilon$ was introduced by \cite{Efstathiou:2012} to model departures from self-similar evolution.  The \cite{Arnaud:2010}  pressure profile with $\epsilon = 0$ provides a  good match to the pressure profiles of massive clusters in the \flamingo\ simulations \citep[see Fig. 3 of][]{Brasspenning:2023}. As can be seen from Fig.~\ref{fig:flamingo}, this simple one-halo model (plotted as the dashed blue line)  gives a very good match to the tSZ power spectrum measured in the \flamingo\ simulation. Notice also that there is no mass bias parameter involved in this comparison because the simulations measure the masses of clusters directly.

Following \cite{Komatsu:2002}, the tSZ power spectrum is given by
\begin{subequations}
\begin{equation}
  D_\ell^{tSZ} = {\ell(\ell+1) \over 2 \pi} \int dz {dV \over dz d \Omega} \int {dn \over dM} dM \vert y_\ell (M, z) \vert^2,
  \label {equ:KS1a}
\end{equation}
where $(dn/dM)dM$ is the halo mass function\footnote{As in \cite{Efstathiou:2012}, we use the \cite{Jenkins:2001} parameterization of the halo mass function.} and in the small angle approximation  $y_\ell$ is given by the following integral over the
pressure profile:
\begin{equation}
  y_\ell = {\sigma_T \over m_e c^2} {4 \pi R_{500} \over \ell^2_{500}} \int dx x^2 {\sin (\ell x/\ell_{500}) \over
    (\ell x/\ell_{500})} P_e(x), 
  \label {equ:KS1b}
\end{equation}
\end{subequations}
where $\ell_{500}$ is the multipole corresponding to the angular size subtended by $R_{500}$ at the  redshift of the cluster.

The contribution to the tSZ power spectrum as a function of multipole,
cluster virial mass and redshift is shown in
Fig.~\ref{fig:evolution}. At $z \simlt 0.5$, 
this figure shows  similar behaviour to Fig.~3 from \citep{McCarthy:2014}, which is
based on the cosmo-OWLS simulations. For the multipoles relevant to \Planck~($\ell \sim 200-500$,  see below)
 shown approximately by the pink shaded bands, the tSZ power spectrum is dominated
by clusters at low redshift ($z \simlt 0.5$) with virial masses $M_V
\simgt 10^{14.5}M_\odot$. M23 argue that baryonic feedback processes in
such massive clusters are unlikely to drastically alter their
pressure profiles. At
the higher multipoles probed by ACT and SPT  ($\ell \sim 2000-3000$)  indicated by the blue bands, the tSZ power spectrum probes  
lower mass clusters  at higher redshifts \citep{Komatsu:2002} and is therefore more
sensitive to feedback processes and departures from
self-similarity. This is illustrated in Fig.~\ref{fig:evolution}, where we have shown
the steep decline in the power spectrum at high redshift if the evolution parameter in Eq.~\ref{equ:P1} is set to  $\epsilon=1$ instead of the self-similar value $\epsilon=0$}. However, even at high multipoles $\sim 3000$, 
M23 conclude that plausible variations in the  \flamingo\ feedback model  cannot reproduce the tSZ
measurements reported by  ACT and SPT assuming the \Planck\ \LCDM\ cosmology.
The main 
purpose of this paper is to critically reassess  the reliability the
tSZ measurements plotted in Fig.~\ref{fig:flamingo}.

\section{Analysis of y-maps}
\label{sec:ymaps}

\begin{figure}
	\centering
	\includegraphics[width=85mm, angle=0]{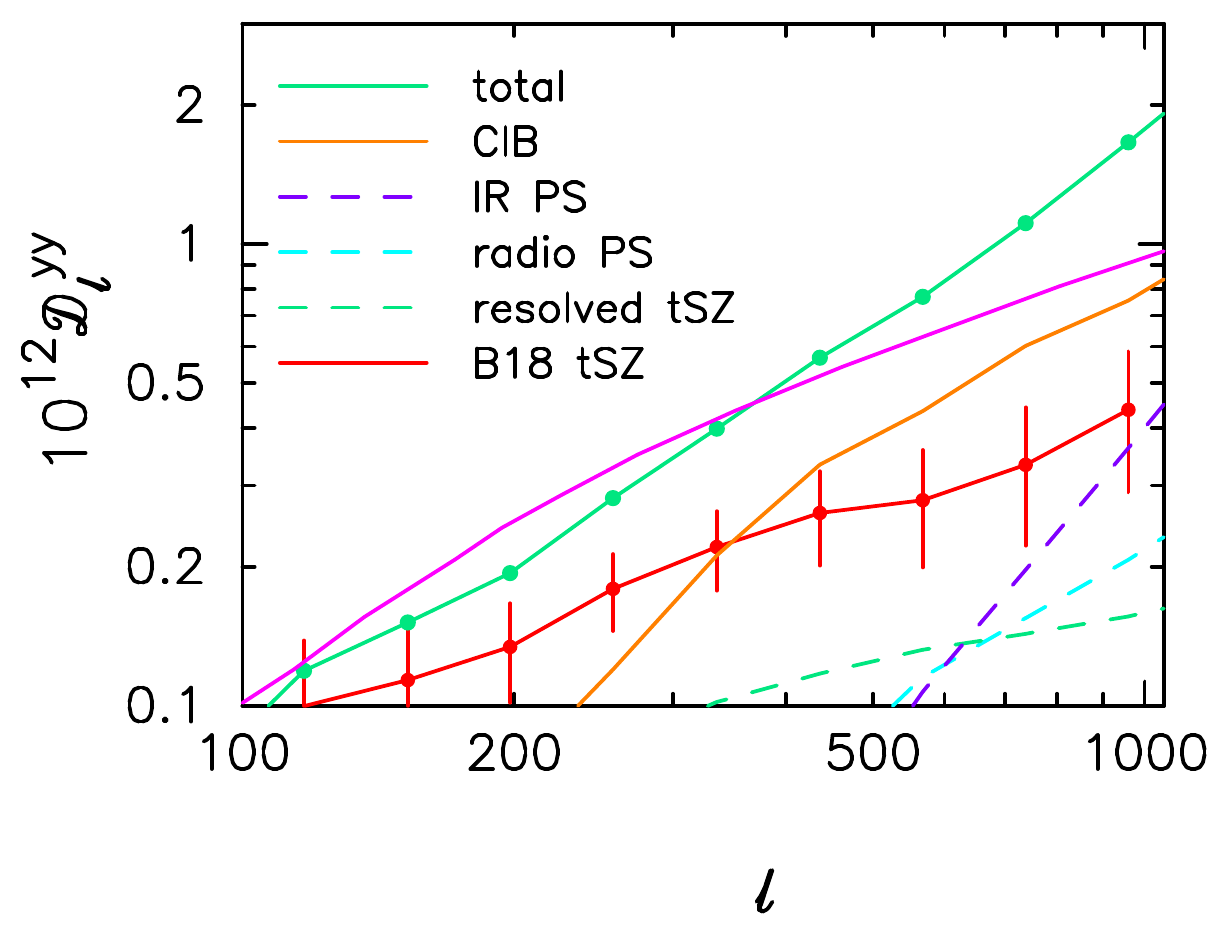} 
	\caption{The green points show the power spectrum of the NILC$\times$MILC y-map cross spectrum analysed by B18.
          The  figure shows the contributions from the clustered CIB, infrared point sources, radio sources, resolved SZ clusters and unresolved tSZ determined by B18 by fitting template power spectra to the green points.   The B18 tSZ power spectrum is a subdominant component of the total power spectrum over most of the multipole range shown in the figure. The purple line shows the tSZ power spectrum from the \flamingo\ simulation of the \LCDM\ cosmology (as plotted in Fig.~\ref{fig:flamingo}).}

	\label{fig:bolliet}

\end{figure}

Figure~\ref{fig:bolliet} shows the decomposition of the
NILC$\times$MILC y-map cross spectrum analysed by B18 into various
components. The total power spectrum is shown by the green points.
The red points show the tSZ power spectrum computed by B18 (as plotted
in Fig.~\ref{fig:flamingo}) after subtraction of the  CIB,
 radio sources and resolved clusters. As can be seen,
the inferred tSZ is a small fraction of the total signal over the
entire multipole range shown in Fig.~\ref{fig:bolliet} and is
therefore extremely sensitive to the assumed shapes of the contaminant
power spectra. B18 adopted template shapes from
\cite{Planck_Ymap:2016} folded through the NILC and MILCA weights.
The clustered CIB component and IR point source amplitudes are from
the models of \cite{Bethermin:2012} and the radio point source
amplitudes are from the models of \cite{Tucci:2011}. The model for the
tSZ spectrum contributed by clusters resolved by \Planck, plotted as the dashed green line in Fig.~\ref{fig:bolliet} is described
in \cite{Planck_Ymap:2016}. There are significant uncertainties
associated with the template spectra. 
Figure \ref{fig:bolliet} implies that the CIB is the
dominant contaminant, yet very
little is known about the amplitude and shape of the CIB power
spectrum at $100$ and $143$ GHz and it is dangerous to rely on the
models of \cite{Bethermin:2012}\footnote{Even at high frequencies of
$350 \mu {\rm m}$ and $500 \mu{\rm m}$ ($860$ amd $600$ GHz) when most
of the CIB is resolved by Herschel \citep{Viero:2013}, the
\cite{Bethermin:2012} models fail at multipoles $\simgt 2000$ (see
\cite{Mak:2017}).}.

\begin{figure}
	\centering
	\includegraphics[width=85mm, angle=0]{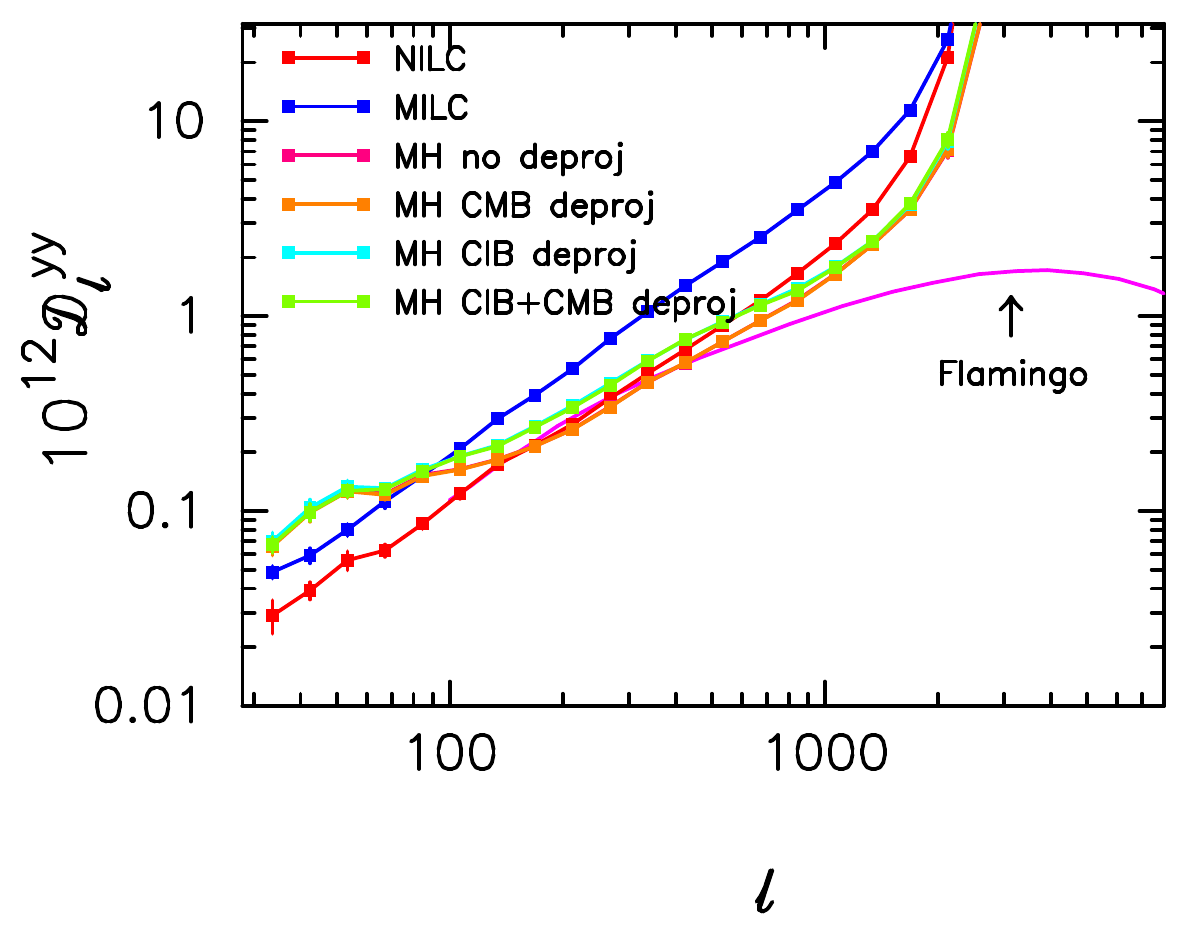} 
	\caption{ The curves labelled NILC and MILC show half-ring cross spectra of the \Planck\  y-maps. The curves labelled MH show half-ring cross spectra of y-maps constructed by 
    \protect\cite{McCarthy:2024} with no deprojection (as in standard NILC) and with additional constraints applied to deproject the CMB,  CIB and CMB+CIB. Note that the pink `MH no deproj' points lie almost exactly under the `MH CMB deproj' points, and the cyan `MH CIB deproj' points lie under the `MH CIB+CMB deproj' points.}

	\label{fig:YYmaps}

\end{figure}

\begin{figure*}
	\includegraphics[width=120mm, angle=0]{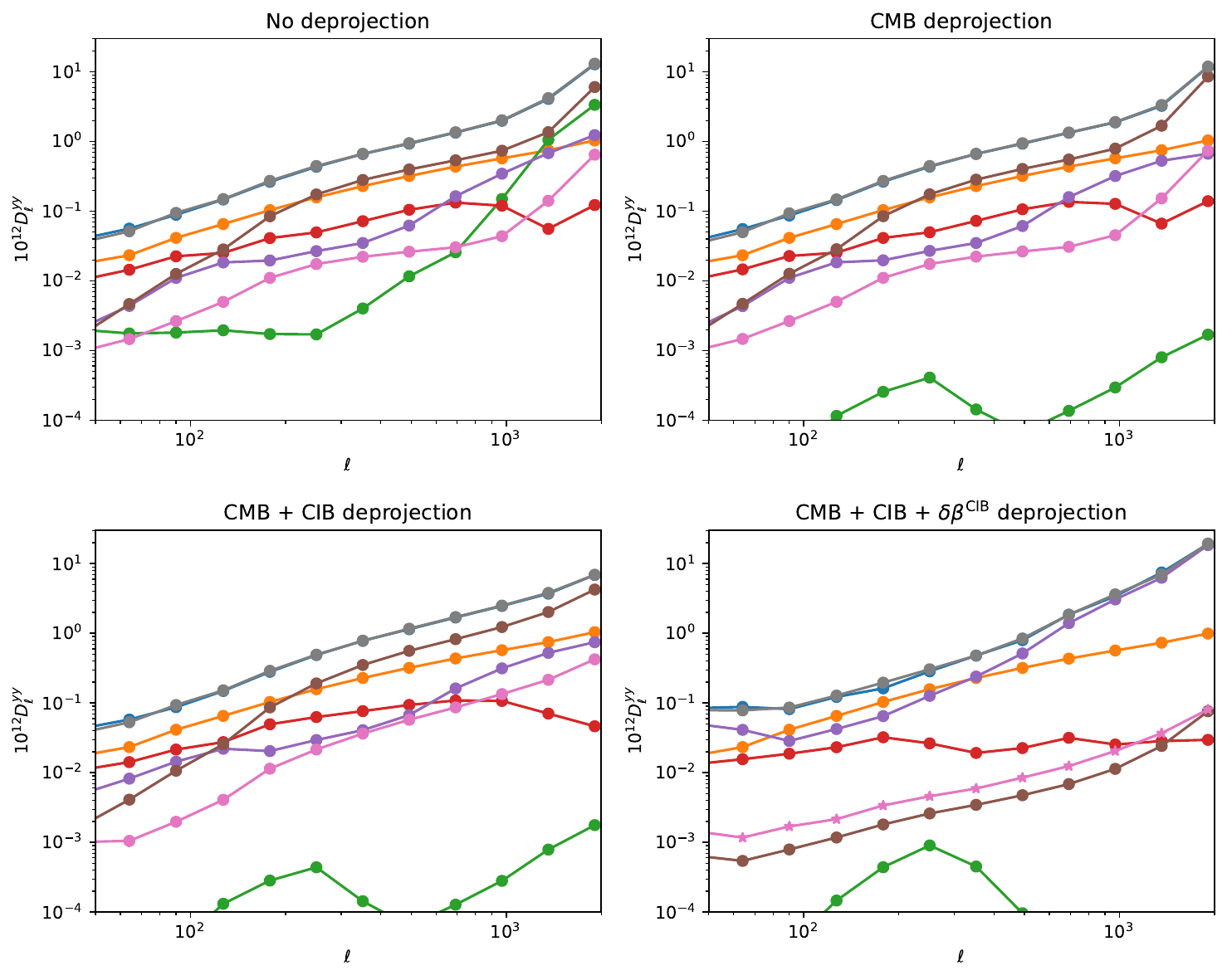} 
	 \raisebox{6cm}{\includegraphics[width=40mm, angle=0]{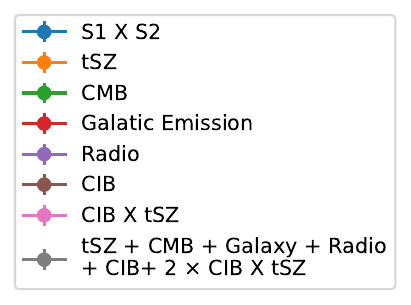} }

    \caption{Contribution of each component to the measured power spectrum of a simulated NILC $y$-map \textit{Planck}-like analysis. The blue points show the measured split power spectrum of our simulations, and the various coloured points show the contribution of the components, with the true tSZ signal in orange, the CIB in brown, and radio point sources in purple. When the CIB is minimized by deprojecting the CIB and its first moment (bottom right), the radio contribution increases to compensate, illustrating the difficulty of simultaneously cleaning all of the foregrounds.}

	\label{fig:ymap_plot}

\end{figure*}

In addition to the y-maps constructed by the \Planck\ collaboration, y-maps have been constructed from the \Planck\
data by other authors  \citep[e.g.][]{Hill:2014,  Tanimura:2021, Chandran:2023,  McCarthy:2024} and also from combinations of
ACT, SPT and Planck maps \citep{Madhavacheril:2020, Bleem:2022,Coulton:2024}.  The blue and red points in Fig.~\ref{fig:YYmaps} show the
NILC and MILC half-ring cross-spectra computed from the y-maps described by \cite{Planck_Ymap:2016}
\footnote{downloaded from https://irsa.ipac.caltech.edu/data/Planck/ \hfill
release$\_$2/all$-$sky$-$maps/ysz$\_$index.html.}. These were computed using the
apodised 70\% sky masks available from the PLA with no corrections for point sources and extended sources. As reported in \cite{Planck_Ymap:2016}, the amplitude of the MILC power spectrum is nearly a factor of two higher than that of the
NILC power spectrum showing that the contamination is sensitive to the map making technique.

The remaining points  in Fig~\ref{fig:YYmaps} show  cross-power
spectra of half-ring split y-maps constructed by \cite[][hereafter MH24]
{McCarthy:2024}\footnote{https://users.flatironinstitute.org/$\sim$mccarthy/ \hfill
ymaps$\_$PR4$\_$McCH23/ymap$\_$standard.}. These were computed using the identical sky masks as those used to compute the
\Planck\ MILC and NILC spectra shown in the figure. MH24 applied a NILC algorithm to the \Planck\ PR4 maps \citep{PR4:2020} but with constraints to deproject
various components. The spectrum labelled `MH no deprojection' shows the results for the standard y-map ILC method (similar to the \Planck\ NILC algorithm),
while the remaining spectra show results for deprojection of the CMB, CIB, and CIB+CMB components respectively. All of these have similar amplitudes. In addition, MH24 applied a moment-based deprojection (based on the work of \cite{Chluba:2017}) which accounts for small variations in the spectral index of the CIB,  though at the expense of increasing the effective noise levels in the reconstructed y-maps.

To gain insight into the contamination of the y-maps, we tested the MH24 algorithms against simulations with known foregrounds. 
The \textit{Planck}-like simulations include extragalactic components from Websky~\citep{Stein:2020its} and Galactic components from PySM3~\citep{Thorne:2016ifb}. The extragalactic components included are the lensed CMB and kSZ (which are included as blackbody components) and the CIB \citep[as described by][]{Stein:2020its} and radio point sources ~\citep[from][]{Li:2021ial}. The simulations were produced at the \textit{Planck} frequencies of 30, 44, 70, 100, 143, 353, and 545 GHz, but  note that the Websky CIB is not provided at frequencies lower than 143 GHz and so for the lower frequency channels we simply rescale the CIB from 143 GHz  using a modified-black-body emission law with dust temperature $20$K and spectral index $1.6$ (see MH24; these parameters are assumed in the deprojection of the CIB). The Galactic components included from PySM3 are thermal dust (\texttt{d1}), synchrotron radiation (\texttt{s1}), anomalous microwave emission (\texttt{a1}), and free-free emission (\texttt{f1}), where the specification in brackets identifies the specific PySM3 model (we refer the reader to the PySM3 documentation for details of these models). At each frequency, we convolve with a \textit{Planck}-like beam and create two realizations of Gaussian white noise (at a level appropriate for the PR4 \textit{Planck} maps), which we add to the simulated sky signal to emulate two independent splits. 

We apply a NILC algorithm similar to that of MH24 to the each set of multifrequency splits  using \texttt{pyilc}\footnote{\url{https://github.com/jcolinhill/pyilc}} (MH24).\footnote{We note that there is a slight difference in the needlet basis used with respect to MH24, although we expect the conclusions to be nearly identical. The needlet basis used in MH24 was a set of Gaussian needlets which followed the \textit{Planck} tSZ NILC analysis (as described in MH24); the simulations described here use a cosine needlet basis.}  We save the computed ILC weights and apply them separately to each of the components, to assess the level of contamination of each component in the final map. We do this for the four deprojection options: the minimum-variance `no deprojection' version; a CMB-deprojected version; and  CMB+CIB and CMB+CIB+$\delta\beta$-deprojected version following the constrained ILC framework described in MH24 \citep[see also][]{Chen:2008gw,Remazeilles:2010hq}. 

By applying the weights separately to each component, we can assess how much of each foreground leaks into the final maps. In Figure~\ref{fig:ymap_plot} we show the measured power spectra of the NILC map and of each component, measured on the area of sky defined by the apodized \textit{Planck} 70\% sky mask. From the `no-deprojection' and `CMB-deprojection' plots (top row of Figure~\ref{fig:ymap_plot}) it is clear that the tSZ contribution (orange lines) only accounts for $\sim 50\%$ of the power spectrum of the full map, with the CIB  (brown lines) the main contaminant, along with a Galactic contribution on the largest scales (red lines). Interestingly, the deprojection of the CIB alone (bottom left row) does not remove a significant amount of CIB power. When we deproject both the CMB and the first moment of the CIB (indicated by CIB+$\delta\beta^{\rm{CIB}}$ deprojection, on the bottom right) we see that the CIB contribution is significantly decreased; however this is at the expense of a compensatory increase in radio source power (purple lines). In this case, the y-map power spectrum is similar to the CMB subtracted 100 GHz power spectrum analysed in the next Section. 

In summary, these simulations demonstrate that y-maps are heavily contaminated by other components and that the nature of the contaminants is sensitive
to the way in which the y-maps are constructed. This is the motivation for seeking another way of extracting the tSZ power spectrum.
\color{black}

\section{The tSZ amplitude inferred from the temperature power spectrum }
\label{sec:tSZspec}

An alternative way of estimating the tSZ effect is to fit a parametric foreground model to  CMB power
spectra measured at several frequencies. This type of analysis has been used to remove foreground contributions from the
\Planck, ACT and SPT temperature power spectra \citep[e.g.][]{Dunkley:2013, Planck_params:2014, Planck_params:2020, Choi:2020, Reichardt:2021}. 
The tSZ amplitude inferred from these investigations, including from \Planck, 
 is consistently lower than the predictions of the \flamingo\ simulations.\footnote{It is for this reason that the \Planck\ cosmological parameter papers used the \cite{Efstathiou:2012} template with $\epsilon=0.5$ (see Eq. \ref{equ:P1}) to flatten the tSZ template compared to the \flamingo\ 
 template of Fig.~\ref{fig:flamingo}.}

 The best fit foreground model \citep[see e.g. Fig. 32 of ][]{Planck_Likelihood:2020} illustrates the difficulty of extracting
 an accurate  tSZ amplitude either from y-maps or from power spectra.
 The tSZ effect in the \Planck\ data  dominates over other foreground contributions only at frequencies of $\sim 100$ GHz
 and only at multipoles $\simlt 500$. At lower frequencies radio sources dominate and at higher frequencies the clustered CIB and Poisson contributions from radio and  infrared sources dominate. In addition, Galactic foregrounds become significant at low multipoles if large areas of sky are used. Power spectrum analyses face similar difficulties to the map-based analyses described in the
 previous section. The tSZ amplitude is small and cannot be extracted without making assumptions concerning the shapes of the
 power spectra of the contaminants, particularly the clustered CIB. However,  there is an advantage in working in the power spectrum domain because one can restrict the range of frequencies to reduce the impact of contaminants with poorly known power spectra.  The goal of this section is to present power spectrum analyses of \Planck,  ACT and SPT that are
 insensitive to the CIB.   We  consider the \Planck\ power spectra in Sect.~\ref{subsec:PlancktSZ}
 and then present  slightly different analyses in  Sect.~\ref{subsec:SPTtSZ}
 tailored to the high multipoles probed by  ACT and  SPT. We present a (template-free) reconstruction of
 the tSZ power spectrum from these experiments in Sect.~\ref{subsec:JointtSZ}.

\subsection{Analysis of Planck spectra}
\label{subsec:PlancktSZ}

\begin{figure}
	\centering
	\includegraphics[width=55mm, angle=-90]{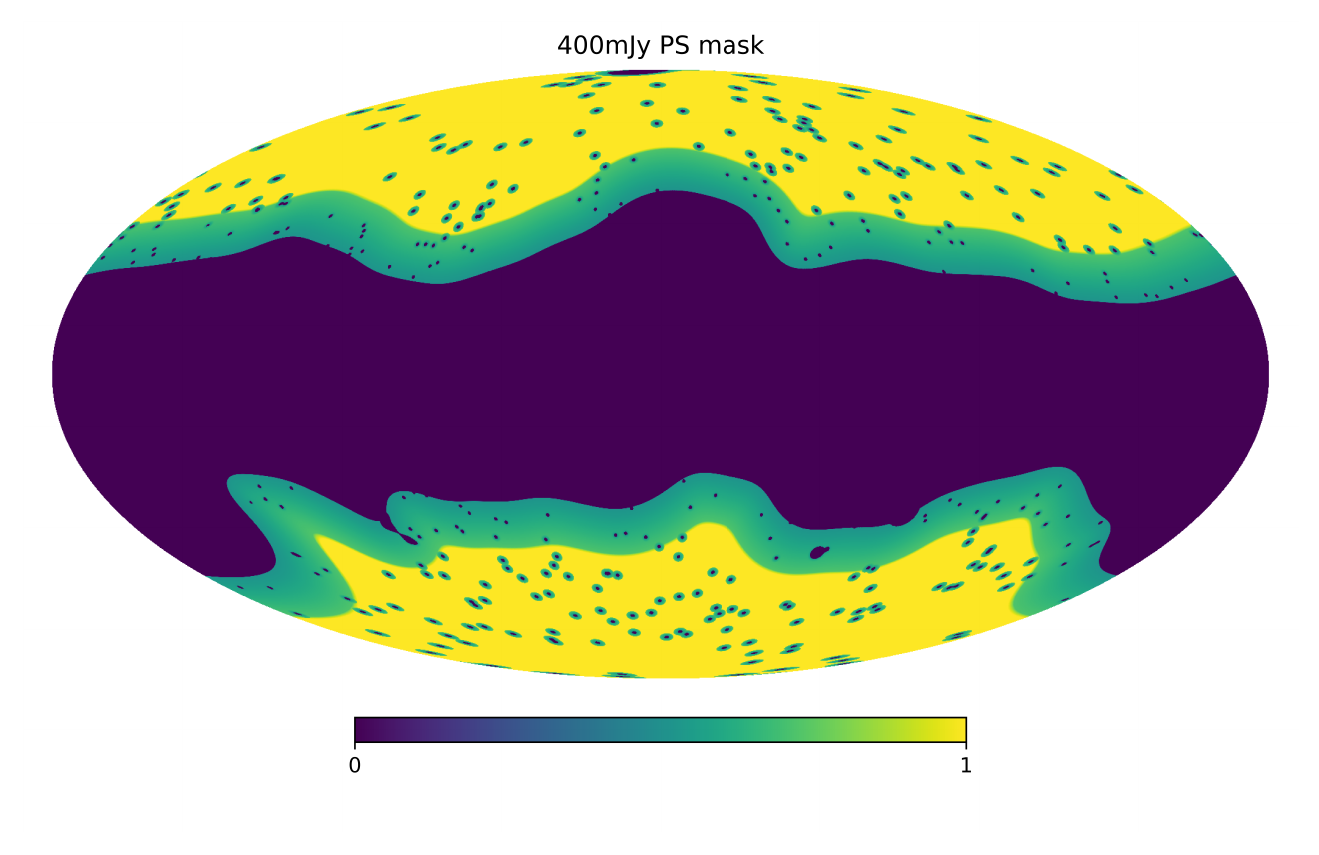} 
	\caption{Mask applied to the Planck maps for the analysis described in Sect.~\ref{subsec:PlancktSZ}.}

	\label{fig:PS_mask}

\end{figure}

The aim of this subsection is to reduce  systematic biases in measurements of the tSZ power spectrum. To achieve this, we
first restrict the sky area that is analyzed  by applying the apodized 50\% sky mask available from the PLA\footnote{Planck Legacy Archive: https://pla.esac.esa.int/\#home.}. This is a smaller sky area than is used for cosmological parameter analysis (see e.g. EG21 who use 80\% sky masks), but is chosen because
in this paper reduction of  biases caused by Galactic emission
is a  more important consideration than increasing the signal-to-noise of the power spectra. In addition to the sky mask,
we mask  sources
with 100 GHz point source flux density (PSFLUX) greater than $400$ mJy listed in the Second \Planck\ Catalogue of Compact Sources
\citep[PCCS2][]{PCCS2:2016}. At this flux limit, the PCCS2 is $\sim 98\%$ complete at 100 GHz (see Fig. 7 of PCCS2). As described below, this high degree of completeness allows us to constrain the Poisson point source  amplitude by  using faint
number counts of radio sources at $100$ GHz.  The point source mask was constructed by applying a sharp symmetric weight function $w_{\rm PS}(\theta)$ as a function of the angular distance $\theta$ relative to the position  of each source,
\begin{equation}
  w_{\rm PS} (\theta) = 1  - e^{-(\theta/\sigma_{\rm PS})^{15}}, \nonumber
\end{equation}
where $\sigma_{\rm PS} = 40^\prime$. To this mask we add the \Planck\ extended object mask and  excise a (lightly apodised)  disc of radius
$2.4^\circ$ centred on the position of the Coma cluster at Galactic coordinates  $\ell =58.6^\circ$, $b= 87.96^\circ$.  The resulting mask
is shown in Fig.~\ref{fig:PS_mask}.

\begin{figure}
	\centering
	\includegraphics[width=85mm, angle=0]{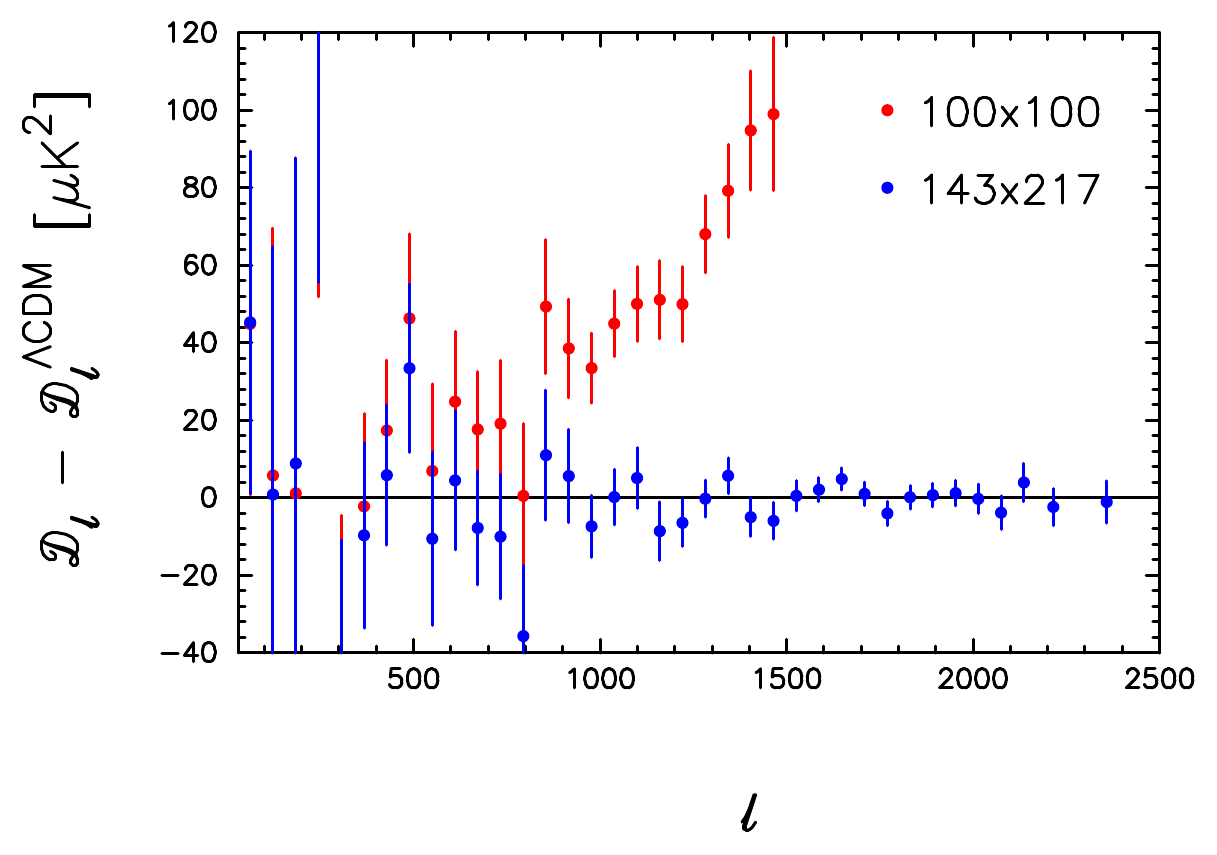} 
	\caption{The $545$ GHz dust-cleaned 100$\times$100 cross spectrum (red points) and dust-cleaned $143\times 217$ cross spectrum (blue points) with the best fit \LCDM\
          spectrum from RGE22 subtracted. The spectra were computed using the 400 mJy 100GHz point source and extended object mask shown in Fig. \ref{fig:PS_mask}. A small correction was applied to the $143\times 217$ cross spectrum (see Eq. \ref{equ:dust_clean}) to remove extragalactic foregrounds at high multipoles. Error bars ($\pm 1 \sigma$) on the bandpowers were computed from the
        \camspec\ covariance matrices.}

	\label{fig:100GHz_compare}

\end{figure}

\begin{figure*}
	\centering
	\includegraphics[width=140mm, angle=0]{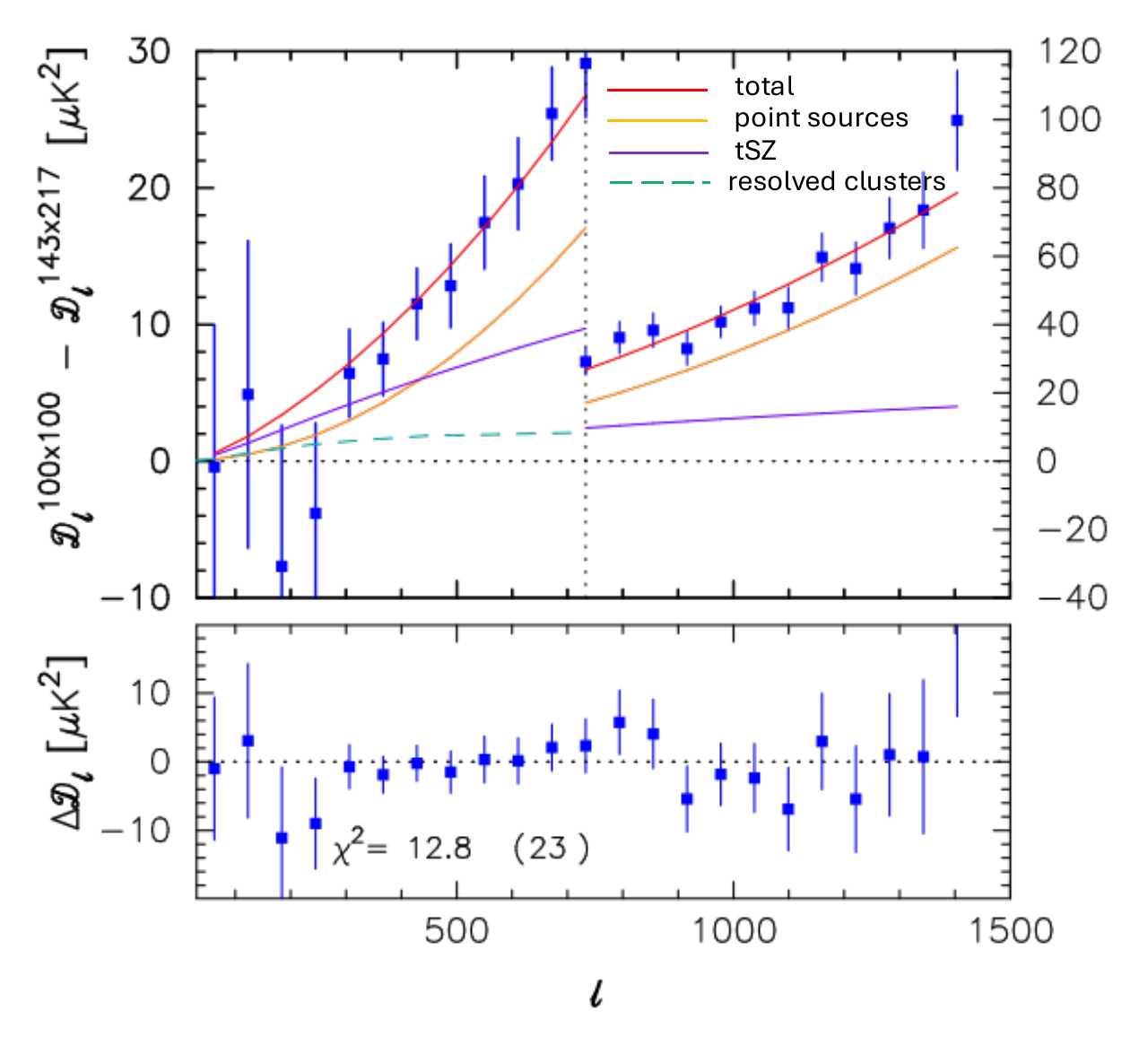} 
	\caption{The upper panel shows the difference of the two spectra plotted in Fig. \ref{fig:100GHz_compare}
          with $\pm 1 \sigma$ errors computed from the bandpower covariance matrix $M_{b b^\prime}$ discussed in the text.
          The tSZ signal from clusters and the Poisson contribution from radio sources are the only significant expected
          contributors to the blue points. The red line in the upper panel show the best fit foreground model which is
          composed of a Poisson radio source component (orange) and a tSZ component (purple) which is modeled as a template with
          the shape of the dashed curve in Fig. \ref{fig:flamingo} with a free amplitude. The green dashed line shows the expected contribution from clusters of galaxies resolved by \Planck\ (from \protect\cite{Planck_Ymap:2016}).   Note that the Coma cluster was masked in our analysis.  Note also the change in the scale of the ordinate in the upper panel at $\ell = 733$. The residuals after subtraction of the foreground model are plotted in the lower panel. We list $\chi^2$ for this fit for  $23$ bandpowers.}

	\label{fig:100GHz_spectrum}

\end{figure*}

We focus on the $100$ GHz power spectrum, since the main contributors at this frequency
are the primary  CMB, tSZ and radio sources\footnote{ 
The CIB model of \cite{Bethermin:2012},  normalized to the best fit  CIB amplitude  at $217$ GHz determined from  a combined CMB+foreground power spectrum analysis to the \Planck\ spectra over the frequency range 100-217 GHz,  has an amplitude of $D_{\ell=500} = 0.25\  \mk2$ (see Fig. 9.2 of \citep{Efstathiou:2021}). This is much smaller that the best fit tSZ amplitude of $D_{\ell=500} \sim 6.9 \mk2$ inferred at 100 GHz (see Fig.~\ref{fig:100GHz_spectrum}). We therefore ignore the CIB contribution at 100 GHz The Planck lower frequencies are dominated by radio sources. In the future, high resolution ground based observations in the frequency range
30-100 GHz \citep{SO_2019} will provide useful information on the tSZ effect.}. 
Throughout this Section, we compute cross spectra from the PR4 Planck A and B  maps \citep{PR4:2020},  following the \camspec\ analysis described by  \cite{Efstathiou:2021} (hereafter EG21) and \cite{Rosenberg:2022} (hereafter  RGE22).  The amplitude of the tSZ contribution to the $100$ GHz power spectrum is expected to be less than $10 \  \mk2$ compared to the amplitude of $\sim 6000 \ \mk2$ at $\ell \sim 200$ of the primary CMB. To detect such a small effect, it is necessary to use the \Planck\ data themselves to estimate the contribution from primary CMB 
in order to eliminate cosmic variance. In our analysis, we subtract the primary CMB using a $545$ GHz
dust-cleaned $143\times217$ GHz cross-spectrum computed using the sky mask
shown  in Fig.~\ref{fig:PS_mask}. The dust cleaning is performed  in the power spectrum domain
as discussed in Sect. 7.3 of EG21. The dust cleaning removes most of the CIB in the $143\times 217$ spectrum  in addition to Galactic dust emission,  leaving a small foreground contribution at high multipoles (see Fig. 11.4 of EG21).  The best fit base TTTEEE \LCDM\ power spectrum of  RGE22 is accurately reproduced by subtracting the following power law  from the dust cleaned $143\times 217$ power spectrum:
\begin{equation} 
  D_\ell^{143\times 217{\rm corr}} =  D_\ell^{143\times 217} - 12.295 (\ell/1500)^{1.701} \mk2.  \label{equ:dust_clean}
\end{equation}
This is close to the best fit foreground model in the fits describd in REG but differs slightly
because REG used different point source masks at $143$ and $217$ GHz.
The residuals of $D_\ell^{143\times217{\rm corr}}$ relative to the best fit \LCDM\ model are shown by the blue points in
Fig.~\ref{fig:100GHz_compare}.

The red points in Fig.~\ref{fig:100GHz_compare} show the resdiuals for the  dust-cleaned $100\times 100$  cross spectrum with no correction for foreground components. At low multipoles $\ell \simlt 500$, the $100\times 100$ and $143 \times 217$ spectra track each other to within $10 - 20 \ \mk2$ because the errors at these low multipoles
are dominated by cosmic variance. At higher multipoles, the
spectra diverge as radio sources become significant in the $100 \times 100$ spectrum.

The difference between these two spectra are shown in the upper panel of Fig.~\ref{fig:100GHz_spectrum}. We have split the figure into two parts so that one can see visually the best fit tSZ contribution at low multipoles. The errors on the difference, $\Delta D_\ell = D^{100\times 100}_\ell  - D^{143 \times 217}_\ell$ are  computed from the \camspec\ covariance matrices which include
small  Gaussian contributions from the best fit foreground model of Eqs. \ref{equ:dust_clean}
and   \ref{equ:tSZ_fit1}.
We also added a trispectrum  contribution  contribution to the covariance matrix (arising from the
angular extent of nearby clusters)  of the bandpowers plotted in Fig. \ref{fig:100GHz_compare}
\begin{subequations}
\begin{equation}
  M^{\rm Tr}_{b b\prime} =  {T_{b b^\prime} \over 4 \pi f_{\rm sky}},  \qquad T_{b b^\prime} = \sum_{\ell \in b} \sum_{\ell^\prime \in b^\prime}
{T_{\ell \ell^\prime} \over N_b N_b^\prime},  \label{equ:Tri1}
\end{equation}
where $N_b$ is the number of multipoles contributing to bandpower $b$, 
\begin{eqnarray}
\hspace{-0.07truein} T_{\ell \ell^\prime} \hspace{-0.07truein}  & = & \hspace{-0.07truein}  {\ell(\ell + 1)\ell^\prime(\ell^\prime + 1)\over 4 \pi^2}   \qquad \qquad \qquad \qquad \qquad   \qquad \qquad  \nonumber \\ 
                   &    &\hspace{-0.08truein} \int dz {dV \over dz d \Omega} \int {dn \over dM} dM \vert y_\ell (M, z) \vert^2  \vert y_\ell^\prime (M, z) \vert^2,  \label{equ:Tri2}
\end{eqnarray}
\end{subequations}
\citep{Komatsu:2002, 2009ApJ...702..368S,2013PhRvD..88f3526H,Bolliet:2018} and $y_\ell(M, z)$ is given in Eq.  \ref{equ:KS1b}. To evaluate this expression we adopt the fiducial tSZ model that was used to produce the dashed curve in Fig.~\ref{fig:flamingo} and set
$f_{\rm sky} =  \sum w_i^2 (\Omega_i/4 \pi) = 0.396$, where the sum extends over all map pixels each of solid angle
$\Omega_i$ and $w_i$ is the weight of the mask at pixel $i$. For the four bandpowers at $\ell \le 300$ plotted in
Fig.~\ref{fig:100GHz_spectrum}, the errors are dominated by uncertainties in the dust cleaning. For these band powers
we replace the elements of the covariance matrix  $M_{b, b}$, $M_{b b+1}$, $M_{b+1, b}$ for $b \le 4$ with the
covariance matrix determined from the scatter of $\Delta D_\ell$ within the bands. The $\pm 1 \sigma$ error bars plotted in Fig.~\ref{fig:100GHz_spectrum} are computed from the diagonals of the final bandpower covariance matrix $M_{b b^\prime}$.

The aim of this analysis is to create a simple linear combination of \Planck\ spectra for which the main contaminant to the tSZ spectrum has a known spectral shape. Having subtracted the primary CMB\footnote{Including the small frequency independent contribution from the kinetic Sunyaev-Zeldovich effect.} and Galactic dust emission, the only significant remaining contributions to the $100 \times 100 - 143 \times 217$ spectrum come from the tSZ effect and Poisson point sources. We model the tSZ effect
using the dashed line of Fig.~\ref{fig:flamingo} as a template multiplied by the parameter $A^{\rm Planck}_{\rm tsz}$. The radio source contribution is modeled as a Poisson spectrum with amplitude
\begin{equation}
   D^{\rm PS}_\ell  = 31.71 A^{\rm Planck}_{\rm PS} {\ell(\ell+1) \over 10^6} \mk2 ,   \label{equ:PS1}
\end{equation}
where the coefficient has been chosen so that  $A_{\rm PS} = 1$ corresponds to the best fit to the $100 \times 100 - 143 \times 217$ spectrum. The relative calibration of the \Planck\ TT spectra is sufficiently accurate that there is no need to sample over calibration parameters (see EG21, Section 9.1.1).

\begin{figure}
	\centering
	\includegraphics[width=85mm, angle=0]{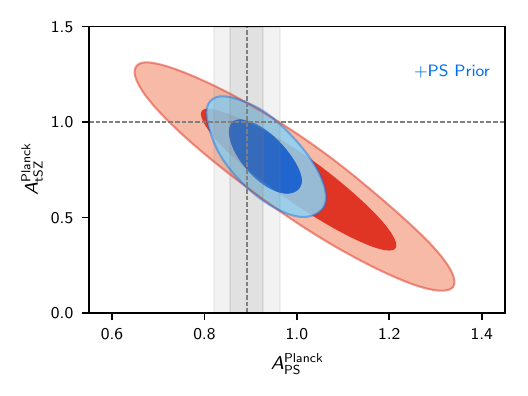} 
	\caption{68\% and 95\% contours on the parameters $A^{\rm Planck}_{tSZ}$ and $A^{\rm Planck}_{\rm PS}$ derived by fitting the
          $100\times 100 - 143\times 217$ power spectrum difference (red contours).  Consistency with the predictions of the \flamingo\
          \LCDM\ prediction of Fig.~\ref{fig:flamingo} requires $A^{\rm Planck}_{tSZ} = 1$ (shown by the dotted horizontal line). The vertical bands show the $1$ and $2\sigma$ constraints on $A^{\rm Planck}_{\rm PS}$ derived from fitting point source number counts at $100$ GHz (see Fig.~\ref{fig:source_counts}).  Blue contours show the results when the number count constraint on $A^{\rm Planck}_{\rm PS}$ is included
          as a prior. }

	\label{fig:SZ_constraint}

\end{figure}

We assume a Gaussian likelihood for the bandpowers with covariance matrix $M_{b b^\prime}$ computed as described above and
 sample over the two free parameters $A^{\rm Planck}_{tSZ}$ and $A^{\rm Planck}_{\rm PS}$ using the 
 {\tt MULTINEST} nested sampler \citep{Feroz:2009, Feroz:2011}.  We find
 \begin{equation}
  \left. \begin{array}{l}
     A^{\rm Planck}_{tSZ}  =  0.706 \pm 0.243,  \\ 
     A^{\rm Planck}_{\rm PS}  =  1.000 \pm 0.140. \end{array} \right \} \label{equ:tSZ_fit1}
 \end{equation}
 These two parameters are highly correlated as illustrated in Fig.~\ref{fig:SZ_constraint}.

 As is evident from Fig.~\ref{fig:100GHz_spectrum}, the unresolved tSZ contribution is a small 
 effect that is difficult to measure accurately from the \Planck\ data. The result of Eq.~\ref{equ:tSZ_fit1} has such a large error that we
 cannot exclude the \flamingo\ prediction of Fig.~\ref{fig:flamingo}. Our results also suggest that the errors on 
 the B18 tSZ power spectrum (and on the power spectra inferred from
 similar analyses of y-maps such as \cite{Tanimura:2021}) have been underestimated because they do not included
 errors in the shapes of the major contaminants.\footnote{For example, \cite{Tanimura:2021} use the \cite{Maniyar:2021} theoretical models which are untested at frequencies below $217$ GHz.}

\begin{figure}
	\centering
	\includegraphics[width=85mm, angle=0]{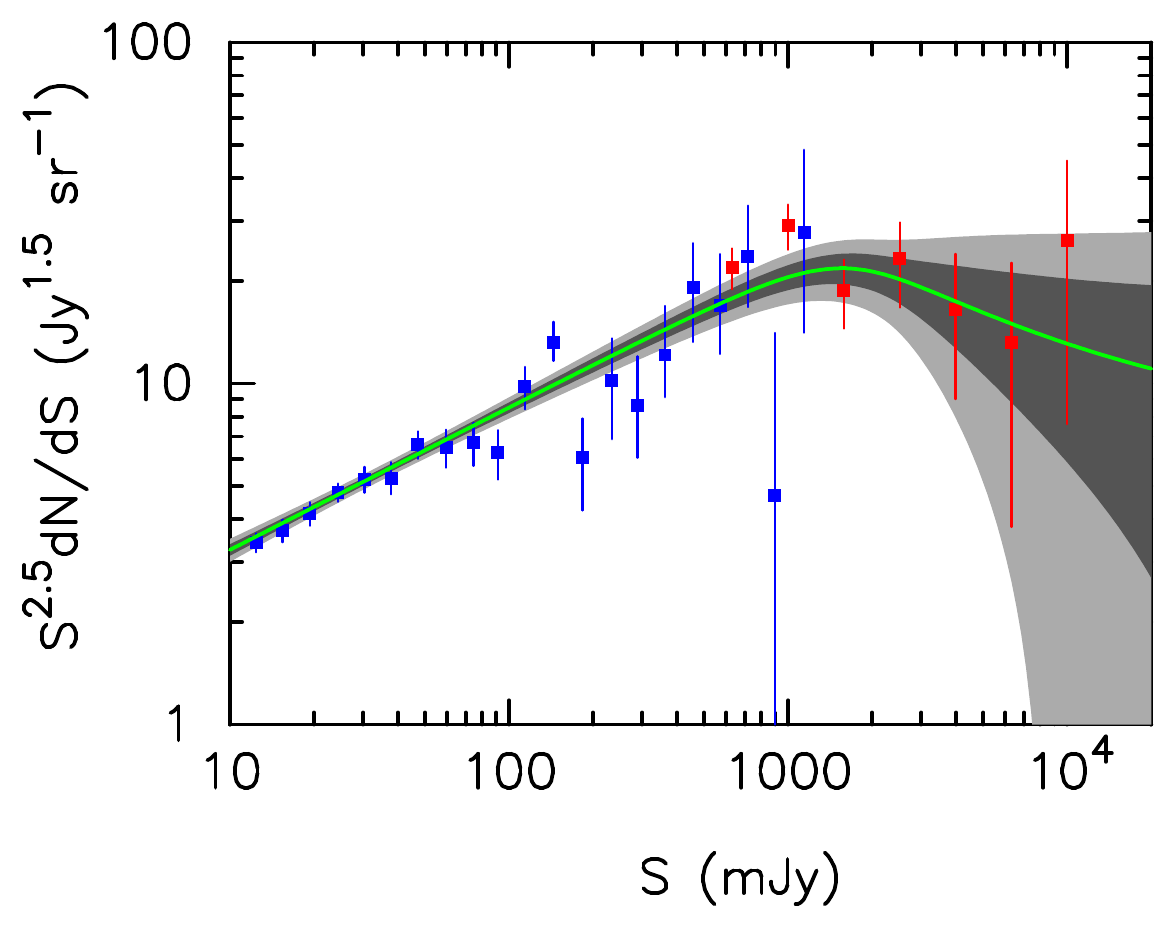} 
	\caption{Source counts at $100$ GHz. The red points show source counts measured from \Planck\ \protect\citep{Planck_PS:2013}.
          The blue points show counts from SPT \protect\citep{Everett:2020} at $95$ GHz rescaled to $100$ GHz. The green line
          shows the best fit to the function of Eq.\ref{equ:PS2} and the grey bands show $1$ and $2\sigma$ errors computed from
          the {\tt MULTINEST} chains.}

	\label{fig:source_counts}

\end{figure}

The degeneracy between $A^{\rm Planck}_{tSZ}$ and $A^{\rm Planck}_{\rm PS}$ can be broken by using source counts at $100$ GHz.  The red points in Fig.~\ref{fig:source_counts} show  $100$ GHz source counts measured by \Planck\ as listed  in Table 7 of \cite{Planck_PS:2013}.  The blue points show the source counts at $95$ GHz from the 2500 square degree SPT-SZ survey \citep{Everett:2020}. We apply a small  correction
to the SPT flux densities in the $95$ GHz band (effective frequency of $93.5$  GHz for a radio source with spectral index $S_\nu \propto \nu^{\alpha_R}$, $\alpha_R \approx -0.5$) to transform to the \Planck\ band frequency at $100$ GHz (effective frequency of $100.84$ GHz\footnote{Interpolating between the numbers given in \cite{Planck_SR:2014} to $\alpha_R = -0.5$.}), giving
$S^{\rm Planck}_{100} = 0.963 S^{\rm SPT}_{95}$.

We fit  the number counts shown in Fig.~\ref{fig:source_counts}  to the  function
\begin{equation}
  S^{2.5} {dN \over dS } = A_c \left ( {x \over 100} \right )^{\alpha_c}
  \left (1 + \left ({x \over x_c} \right )^{\beta_c} \right)^{\gamma_c}, \quad x = 1000S, \label{equ:PS2}
\end{equation}
using {\tt MULTINEST}. The marginalized posteriors  of the parameters are found to be
\begin{equation}
  \left. \begin{array}{l} A_c  =  \ 8.55 \  (8.51) \pm 0.35 \  {\rm Jy}^{1.5}{\rm sr}^{-1}, \\
  x_c  =  1565 \ (1101)\  \pm 420,  \\
 \alpha_c  =  \ 0.419 \  (0.421) \pm 0.025,  \\
 \beta_c   =   \ 3.63\ (6.73)\ \pm 1.65,    \\
 \gamma_c  =   \ 0.307 \ (0.098) \pm 0.177,  
  \end{array} \right \}  \label{equ:PS3}
\end{equation}
where the numbers in brackets give the best fit values of the parameters. The best fit and $\pm 1 \sigma$ error bars computed from the {\tt MULTINEST} chains are plotted  in Fig.~\ref{fig:source_counts}.

The power spectrum of Poisson distributed point sources is given by
\begin{equation}
  C^{\rm PS}_\ell = \int_0^{S_{\rm lim}} S^2 {dN \over dS } dS. \label{equ:PS}
\end{equation}
Applying the monochromatic conversion from Jy to thermodynamic temperature, 
\begin{equation}
\Delta T = {(e^x - 1)^2 \over x^2 e^x} {c^2 I_\nu \over 2 \nu^2 k},  \quad x = {h\nu \over kT}, \label{equ:PS4}
\end{equation}
the point source amplitude at $\ell = 1000$  at  $100$ GHz
in temperature units is given by 
\begin{equation}
  D^{\rm PS}_{1000} =  (0.00413)^2 {10^6 \over 2 \pi}  \int_0^{S_{\rm lim}} S^2 {dN \over dS } dS \   \mk2.  \label{equ:PS5}
\end{equation}
We evaluate this integral for $S_{\rm lim} = 400 \ {\rm mJy}$ and monitor  $D^{\rm PS}_{1000}$ as a derived parameter in the {\tt MULTINEST} chains. The results give
\begin{equation}
  D^{\rm PS}_{1000} =  29.2 \pm 1.8  \ \mk2,  \label{equ:PS5}
\end{equation}
which is reassuringly close to the best fit value of Eq.~\ref{equ:PS1} determined by fitting to the power spectrum.
The point source amplitude determined from the number counts breaks the degeneracy between $A^{\rm Planck}_{\rm tSZ}$ and $A^{\rm Planck}_{\rm PS}$ as
shown  in  Fig.~\ref{fig:SZ_constraint} and favours values of $A_{\rm tSZ}$ close to unity. This is illustrated by the blue contours in Fig.~\ref{fig:SZ_constraint} in which we have imposed the number count constraint of Eq.~\ref{equ:PS5} as a prior on $A^{\rm Planck}_{\rm PS}$. In this case we find
 \begin{equation}
  \left. \begin{array}{l}
     A^{\rm Planck}_{tSZ}  =  0.815 \pm 0.128,  \\ 
     A^{\rm Planck}_{\rm PS}  =  0.931 \pm 0.052. \end{array} \right \} \ {\rm including} \ {\rm PS} \ {\rm prior}.  \label{equ:tSZ_fit2}
 \end{equation}

In summary, we have focussed on the $100$ GHz \Planck\ band. At this frequency,
the power spectrum of radio sources, which has
a known spectral shape, is the main contaminant to the tSZ signal after subtraction of the primary CMB. Our main conclusion, evident from  Fig.~\ref{fig:SZ_constraint}, is that it is difficult
to make an accurate measurement of the tSZ amplitude  from   \Planck\ even if we
apply the  point source prior of Eq.~\ref{equ:PS5}.  the constraint on
     $A^{\rm Planck}_{tSZ}$ cannot exclude the 
\flamingo\  \LCDM\ prediction shown in Fig~\ref{fig:flamingo}. It must be noted, however, that most of the statistical weight
in Eq.~\ref{equ:tSZ_fit2} comes from multipoles $\ell \sim 300-500$. \Planck\ has  little sensitivity to the tSZ spectrum at
higher multipoles. This will become clearer in Sec.
\ref{subsec:JointtSZ} where we present the results of a template-free tSZ power spectrum reconstruction.

\subsection{Analysis of ACT and SPT spectra}
\label{subsec:SPTtSZ}

Figure \ref{fig:flamingo}  shows a large discrepancy between the predictions of the \flamingo\ \LCDM\
tSZ spectrum and the  amplitude inferred from ACT and  SPT at high multipoles. As mentioned above,
the ACT and SPT constraints are derived by fitting a parametric model to power spectra over the frequency
range $\sim 95 - 220$ GHz. These models include templates for a number of foreground components including
the clustered CIB, which  we have emphasised,  
is poorly known at these frequencies. In this section, we focus attention on the power spectra measured
from the SPT-SZ and SPTpol surveys reported by \cite{Reichardt:2021} (hearafter R21) and the ACT deep surveys reported by
\cite{Choi:2020} (hereafter C20). As in the previous section, our
aim is to simplify the analysis so that the inferred tSZ power spectrum is insensitive to the CIB. We therefore
restrict the analysis to R21 95 GHz and C20 98 GHz spectra (thus excluding the  R21 150 and 220 GHz and C20 150 GHz spectra). As in the \Planck\ analysis,  the tSZ effect
has the largest contrast relative  to other foreground components in the ACT and SPT spectra  at these frequencies \footnote{C20
also analyse data from a wide field survey. We do not use the wide data here because the point source contribution
to the power spectrum at $98$ GHz has a much higher amplitude compared to the deep survey. The wide survey is therefore
much less sensitive to the tSZ effect compared to the deep survey.} (see e.g. Fig. 2 of R21).

\begin{figure}
	\centering
	\includegraphics[width=85mm, angle=0]{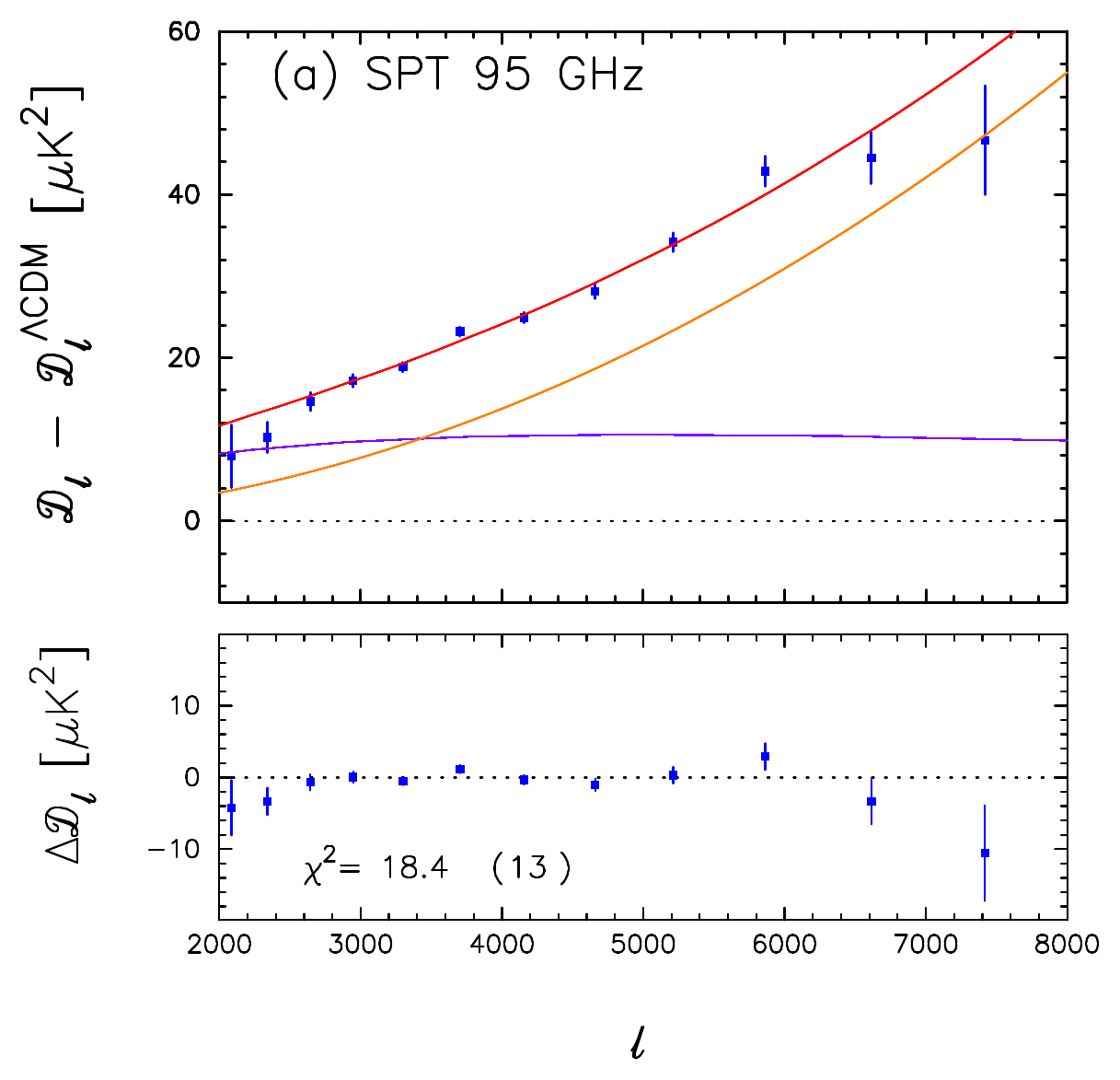} \includegraphics[width=85mm, angle=0]{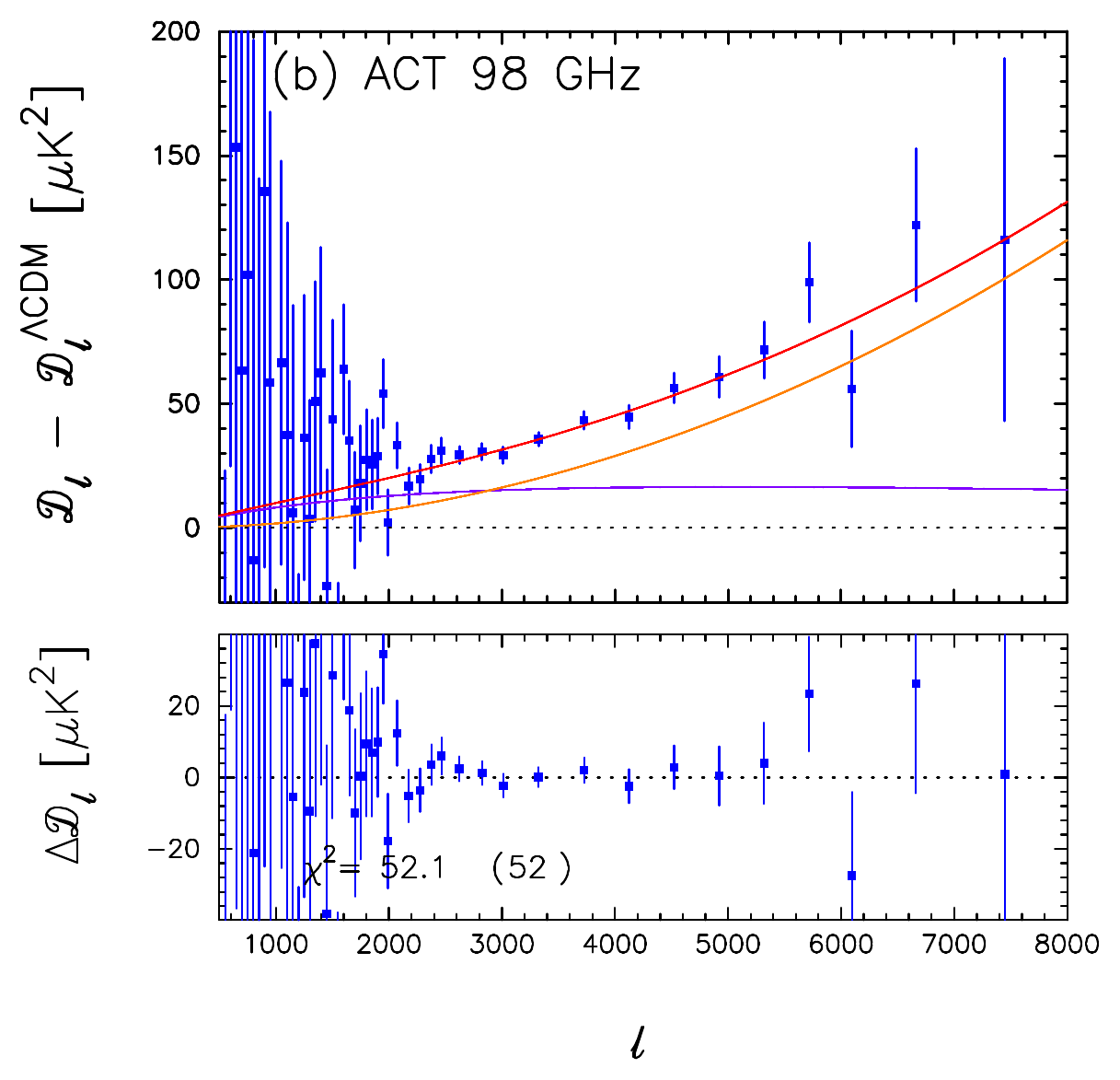} 
	\caption{The upper panels in each plot show the differences between the 95 GHz SPT and 98 GHz ACT
          bandpowers and the power spectrum of
          the best fit \LCDM\ cosmology from RGE21. The $\pm 1 \sigma$ errors on the bandpowers were computed from the diagonals of the SPT and ACT covariance matrices.  The lines show the best fit foreground model.
          The total foreground is shown in red, tSZ
          contribution is shown in purple, and the Poisson point source contribution is shown in orange.
          The residuals after subtraction of the foreground model are plotted in the lower panels. We list  $\chi^2$ for the
          best fits for  $13$ SPT bandpowers and $52$ ACT bandpowers.}

	\label{fig:95GHz_spectrum}

\end{figure}

We use the public releases of the R21 and C20 bandpowers, window functions, beam and bandpower  covariance matrices\footnote{
Downloaded from \\
http://pole.uchicago.edu/public/data/reichardt20/ \\
https://lambda.gsfc.nasa.gov/product/act/act$\_$dr4$\_$likelihood$\_$get.html} and fit the bandpowers to a model consisting of
the best fit \LCDM\ power spectrum from RGE22, the \flamingo\  tSZ template (dashed line in Fig.~\ref{fig:flamingo}) with  free amplitudes $A^{\rm SPT}_{\rm tSZ}$ $A^{\rm ACT}_{\rm tSZ}$, and  Poisson point source components with amplitudes
\begin{equation}
D_\ell^{\rm PS} = \left\{ \begin{array}{l l} \ \ 7.71 A^{\rm SPT}_{\rm PS} \ell(\ell+1)/ 9 \times 10^{-6} \ \mk2,  &  {\rm SPT}, \\
      16.25 A^{\rm ACT}_{\rm PS} \ell(\ell+1)/ 9 \times 10^{-6} \ \mk2,   &  {\rm ACT}  .
      \end{array} \right. \label{equ:SPT}
\end{equation}
The coefficients in Eq. \ref{equ:SPT} are chosen so that $A^{\rm SPT}_{\rm PS}$ and $A^{\rm ACT}_{PS}$ are close to unity
for the best fits described below. 
We allow relative calibration coefficients $c^{\rm SPT}$ and $c^{\rm ACT}$ between \Planck\ and SPT and ACT spectra, such that
$c^{\rm SPT/ACT} D^{\rm SPT/ACT}_\ell = D^{\rm Planck}_\ell$, which we include in the likelihood by imposing Gaussian priors on
$c^{\rm SPT}$ and $c^{\rm ACT}$ with  means of unity and  dispersions of $0.6\%$ (SPT) and $1\%$ (ACT) \footnote{Note that relative calibration of \Planck\ and SPT
at the map level leads to an uncertainty of $0.33\%$ in power  (see Sect. 2.2 of R21) and  to an uncertainty of
$1\%$ in power for ACT (see Sect. 7.1 of C20).}.  We form likelihoods
as described in R21 and C20  and use {\tt MULTINEST} to sample over the free parameters. We find
 \begin{equation}
     \left. \begin{array}{l}
     c^{\rm SPT} = 1.0057 \pm 0.0054,  \\
     A^{\rm SPT}_{\rm tSZ}  =  0.297 \pm 0.023,   \\
     A^{\rm SPT}_{\rm PS}  =  1.000 \pm 0.051,  
   \end{array} \right \}  \label{equ:SPT_fit}
 \end{equation} 
 and
 \begin{equation}
\left.    \begin{array}{l}
     c^{\rm ACT} = 0.9918 \pm 0.0082,  \\
     A^{\rm ACT}_{\rm tSZ}  =  0.463 \pm 0.096,   \\
     A^{\rm ACT}_{\rm PS}  =  1.003 \pm 0.139.  \end{array} \right \} \label{equ:ACT_fit}
 \end{equation}
The differences between the SPT and ACT and power spectra and the \Planck\ best fit model are shown in the
 upper panels of each of Figs.~\ref{fig:95GHz_spectrum}(a, b)  together with the best fit foreground model. The residuals
 with respect to the best fit foreground model are shown in the lower panels.
 The low values of $A^{\rm SPT}_{\rm tSZ}$ and $A^{\rm ACT}_{\rm tSZ}$  are particularly striking because they exclude the \flamingo\ \LCDM\ model
 at very high significance. These results are qualitatively consistent with the estimates of the tSZ amplitudes from SPT and ACT plotted in Fig.~\ref{fig:flamingo}.

We note the following points:

\smallskip

\noindent
(i) We have neglected the kinetic Sunyaev-Zeldovich (kSZ)
  effect. The analysis of multifrequency power spectra show that it is
  a small effect \citep[e.g.][]{Reichardt:2012, Planck_params:2020,
    Choi:2020, Reichardt:2021} with an amplitude that is highly model dependent.
  For example,  \cite{Reichardt:2012} in their analysis of two years of observation with SPT, derived the joint constraint:
  \begin{subequations}
  \begin{equation}
    D^{tSZ_{150}}_{3000} + 0.5D^{kSZ_{150}}_{3000} = 4.60 \pm  0.63 \ \mk2 ,  \label{equ:kSZ1}
  \end{equation}
  for the amplitudes at $\ell=3000$  of the tSZ and kSZ power spectra measured at $150$ GHz. \cite{Choi:2020}
  find
  \begin{equation}
    D^{tSZ_{150}}_{3000} = 5.29\pm 0.66\ \mk2, \ D^{kSZ_{150}}_{3000} < 1.8 \ \mk2 (95\%),  \label{equ:kSZ2}
  \end{equation}
  while \cite{Reichardt:2021} find
  \begin{equation}
    D^{tSZ_{150}}_{3000} = 3.42\pm 0.54 \mk2, \ D^{kSZ_{150}}_{3000} =  3.0 \pm 1.0  \mk2 ,   \label{equ:kSZ3}
  \end{equation}
    \end{subequations}
  and that the tSZ and kSZ amplitudes 
  are correlated as a consequence of the tSZ-CIB cross-correlation (which is very poorly known, see e.g. \cite{Addison:2012}). The correlation from Fig.~3 of R21 is well approximated by 
  $D^{tSZ_{150}}_{3000} + 0.5D^{kSZ_{150}}_{3000} \approx  5 \ \mk2$, consistent with Eq.~\ref{equ:kSZ1}. We will refer to the results in Eqs. \ref{equ:kSZ2} and \ref{equ:kSZ3} as the ACT and SPT SZ measurements respectively. 

  \smallskip
  
  \noindent
  (ii) The tSZ amplitudes of Eq.~\ref{equ:SPT_fit}  and \ref{equ:ACT_fit} correspond to amplitudes at $95$ and $98$ GHz
  of $D^{SPT_{95}}_{3000} = 10.39\  \mk2$ and $D^{ACT_{98}}_{3000} = 15.3 \ \mk2$. Converting the tSZ amplitudes at $150$ GHz quoted in (i),
  we find $D^{tSZ_{98}}_{3000} = 14.26 \pm 1.8 \ \mk2$(ACT)  and $D^{tSZ_{95}}_{3000} = 9.1 \pm 1.4 \ \mk2$ (SPT).  The kSZ contribution is frequency independent and,  as noted above,  is extremely uncertain. In our analysis, we have neglected the kSZ effect, and so our results
  could overestimate the amplitude of the tSZ effect
  by up to a few $\mk2$. However, it is clear from Fig.~\ref{fig:95GHz_spectrum} that the \flamingo\ tSZ template, which predicts $D^{tSZ_{100}}_{3000} \approx 32 \ \mk2$, is firmly excluded and cannot be reconciled with the
  data by any plausible changes to the primary CMB and foreground model. The \LCDM\ \flamingo\ simulations are therefore strongly
  discrepant with observations of the tSZ effect at high multipoles.

  \smallskip

  \noindent
  (iii) In Sect. \ref{subsec:PlancktSZ} we applied a prior based on point source number counts to
  reduce the degeneracy  between $A_{\rm tSZ}$ and $A_{\rm PS}$. Both ACT and SPT mask point sources
  identified at $150$ GHz and so it is not possible to use  source counts to predict the point source
 power at $\sim 100$ GHz without separating infrared galaxies
 from radio sources and making
 assumptions about the spectral indices of the sources. Fortunately, the tSZ amplitudes from ACT and SPT
 are tightly constrained without application of an external constraint on the  point source amplitude.

  \smallskip
  
  \noindent
  (iv) The amplitude of the tSZ template inferred from \Planck,  which is weighted towards multipoles of $\sim 300-500$ (Fig.~\ref{fig:100GHz_spectrum}),  is
  $A_{\rm tSZ} \sim 0.8$. For ACT, which is weighted to multipoles of $\sim 2000 - 2500$ (Fig.~\ref{fig:95GHz_spectrum}b)
  we find $A_{\rm tSZ} \sim 0.46$. For SPT  which is weighted to multipoles of $\sim 2500 - 3500$ (Fig.~\ref{fig:95GHz_spectrum}b), 
  we find $A_{\rm tSZ} \sim 0.297$. These results show a trend for $A_{\rm tSZ}$ to decrease as we probe higher multipoles,  suggesting that the true tSZ power spectrum may be shallower than the \flamingo\ template used to derive these numbers. We explore this possibility in the next subsection.

\begin{figure}
	\centering
	\includegraphics[width=85mm, angle=0]{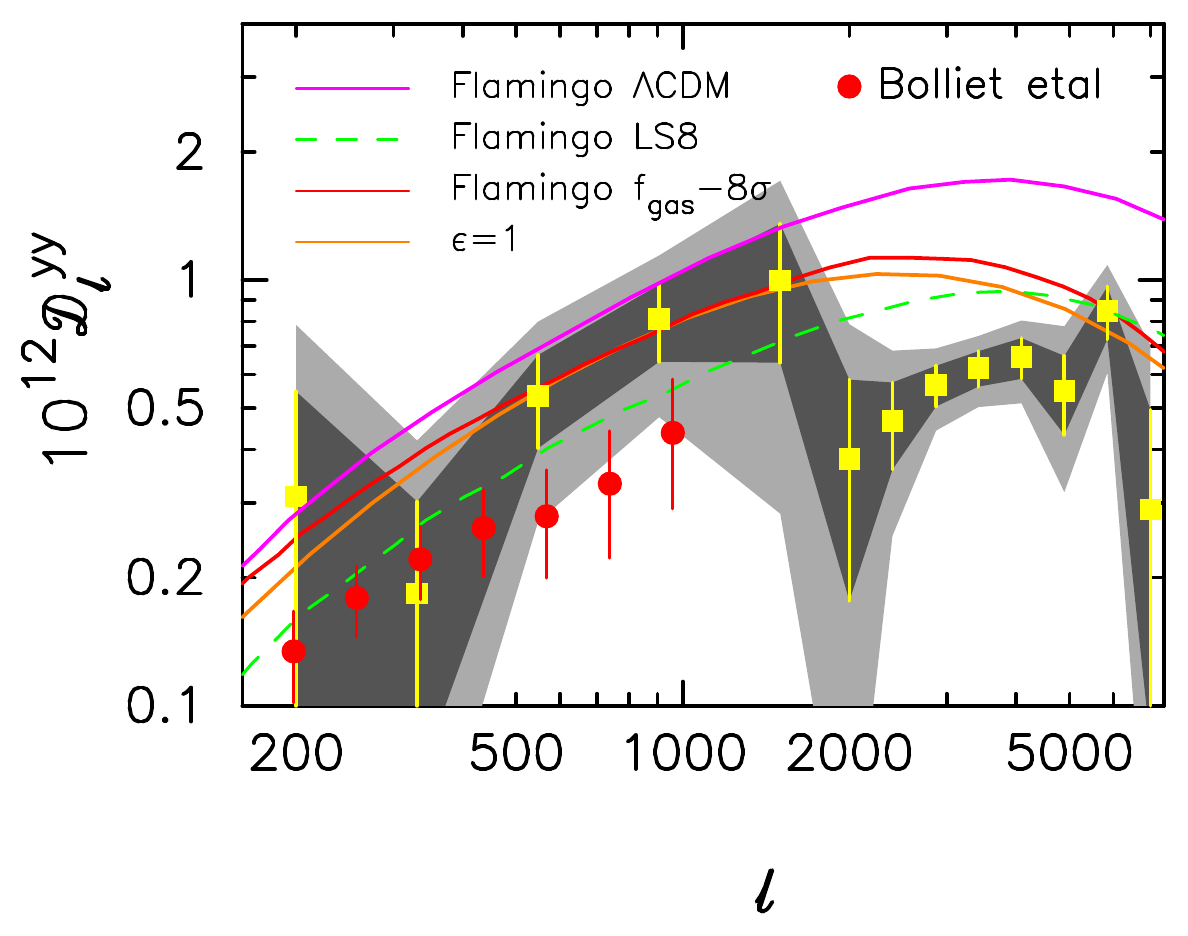} 
	\caption{Reconstruction of the tSZ power spectrum derived by combining the \Planck, ACT and SPT likelihoods of the
          previous subsections (yellow points). We solve for the amplitude of $D^{yy}$ at each of $13$ node points and interpolate the tSZ spectrum between the nodes shown in the Figure. The shaded bands show the $1$ and $2 \sigma$ errors. The red points show the tSZ spectrum inferred by B18 as
          plotted in Fig.~\ref{fig:bolliet}. The curves show the baseline \LCDM\
          \flamingo\ prediction from Fig.~\ref{fig:flamingo} (purple line), results  for the \flamingo\ LS8 (low $S_8$) model (dashed green line), a \flamingo\ simulation with enhanced baryonic feedback (red line labeled $f_{\rm gas} - 8\sigma$, see text) 
          and the halo model of Sect.~\ref{sec:motivation}  with evolution parameter
          $\epsilon = 1$ (orange line). The violet dashed line shows the best fit template tSZ spectrum deduced from  ACT-DR6 (see Sec.~\ref{sec:conclusions})}

	\label{fig:multinest}

\end{figure}

\subsection{Template free analysis}
\label{subsec:JointtSZ}

In this subsection we combine the \Planck, ACT and SPT likelihoods described above and  solve for the shape
of the tSZ power spectrum neglecting any contribution from the kSZ effect. The amplitudes of the spectrum $D^{yy}_{\ell_{\rm node}}$ at a set of node points $\ell_{\rm node}$
are treated as free parameters. The tSZ spectrum in between node points is computed by linear interpolation in $\log_{10} \ell$.
The node points are specified in Table \ref{tab:yyspec}. We then run {\tt MULTINEST} to solve for the $13$ amplitudes $D^{yy}_{\ell_{\rm node}}$, $3$ point source amplitudes $A^{\rm Planck}_{\rm PS}$, $A^{\rm ACT}_{\rm PS}$, $A^{\rm SPT}_{\rm PS}$, with a
number count prior on $A^{\rm Planck}_{\rm PS}$ as described in Sect.~\ref{subsec:PlancktSZ}, 
and two calibration parameters $c^{\rm ACT}$ and $c^{\rm SPT}$ with Gaussian priors as discussed in
 Sect.~\ref{subsec:SPTtSZ}.

 The results are summarized in Table \ref{tab:yyspec} and in Fig.~\ref{fig:multinest}. The constraints from \Planck\ are tightest
 at $\ell \sim 500$ and flare out at lower and higher multipoles. The reconstructed power spectrum shows a dip at  $\ell \sim  2000$ which comes from the lowest two band powers in the SPT spectrum plotted in Fig.~\ref{fig:95GHz_spectrum}a. The best fit to the ACT 98 GHz spectrum actually shows a small excess at $\ell \sim 2000$ (see Fig.~\ref{fig:95GHz_spectrum}b) but the ACT spectra contribute relatively low statistical weight compared to \Planck\ and SPT. The results appear to show a jump in power at $\ell \sim 2000$, but it is important to recognise that the \Planck\ points are strongly correlated and can move in lockstep towards the top or bottom of the error ranges depending on the amplitude of the radio point source amplitude. We also show the points from B18, which lie at the bottom end of the $\sim 2\sigma$ error range for our measurements. Continuity with the results from ACT and SPT suggests that the true amplitude of the tSZ spectrum at $\ell \simlt 1000$ lies at the lower
 end of the error range. Overall,
 the results shown in Fig.~\ref{fig:multinest} show a large discrepancy at high multipoles with the baseline \LCDM\ \flamingo\ prediction  at
 $\ell \simgt 2000$. In addition, the amplitude $D^{yy}$ inferred from Planck at $\ell \sim 500$  is similar to the amplitude inferred at
 $\ell \sim 3000$, thus the tSZ spectrum must have a shallower slope than the baseline \LCDM\ \flamingo\ prediction. 
 We defer further interpretation of these results to the next Section.

 \begin{table}

\begin{center}

  \caption{Reconstruction of the tSZ power spectrum using \Planck, ACT and SPT power spectra.
    The first column gives the value of the  multipole at each of $13$ nodes. The second column
    gives the estimate of the $yy$ power spectrum at each node point. The tSZ spectrum is interpolated linearly in $\log_{10}\ell$  between these nodes. The third column gives the $1 \sigma$ error on $10^{12} D^{yy}_{\ell_{\rm node}}$. }

\label{tab:yyspec}

\smallskip

\begin{tabular}{|c|c|c|} \hline
  $\ell_{\rm node}$ & $10^{12}D^{yy}_{\ell_{\rm node}}$ &  $1 \sigma$ error\\
  200.  & 0.310 &  0.237 \\
 330.97 & 0.184 &  0.118 \\
547.72  & 0.535 &  0.132 \\
 906.4 &  0.810 &  0.166 \\
 1500. &  0.997 &  0.357 \\
 2000. & 0.380 &   0.204 \\
2391.96 & 0.467 &  0.108 \\
2869.74 & 0.567 &   0.0619\\
 3421.38 &  0.621 &  0.0591\\
 4091.91 & 0.659 &   0.073 \\
 4893.84 & 0.548 &   0.114 \\
 5852.94 & 0.846   &  0.120 \\
 7000. &   0.289 & 0.207 \\ \hline
\end{tabular}
\end{center}
\end{table}

 \section{Discussion and Conclusions}
 \label{sec:conclusions}
 
 The aim of this paper has been to present an alternative (and transparent) 
way of measuring the tSZ power spectrum compared to the usual
approach based on y-maps. As discussed in Sect.~\ref{sec:ymaps}, all y-maps
are contaminated by other components and require assumptions concerning the shapes of their power spectra to extract a tSZ power spectrum.

 In this paper, we have concentrated on fitting temperature power spectra at 100 GHz, where the dominant contributions come from the primary CMB, tSZ and radio point sources. The latter component can be modeled  
 accurately by  a Poisson power spectrum $D^{PS}_\ell \propto \ell^2$. We do not consider higher frequencies since they require an accurate model of the CIB, and also the cross-correlation of the tSZ signal with the CIB,  in order to extract the subdominant  tSZ signal.
 
 The tSZ power spectrum that we infer from \Planck\ is consistent with those inferred from \Planck\ y-maps (B18, \cite{Tanimura:2021})  but has larger errors,  which 
 we believe are  more realistic. As a consequence, our analysis of \Planck\ cannot exclude the  \flamingo\ \LCDM\ tSZ
  spectrum.

  However, a similar analysis applied to the ACT 98 GHz and SPT 95 GHz provides convincing evidence of a
  large discrepancy with the \flamingo\ model at multipoles $\ell \simgt 2000$.  The results from
  ACT and SPT spectra  are consistent with each other and also with earlier analyses of ACT and SPT
  \citep{Reichardt:2012, Dunkley:2013}. The low amplitude of the tSZ spectrum at high multipoles is 
  therefore a robust result and must be reproduced in cosmological hydrodynamical simulations that claim to
  match reality. We consider the following two possibilities to explain the discrepancy:

   \begin{table*}

\begin{center}
    
  \caption{Measurements of $S_8$ assuming the base \LCDM\ cosmology. }

\label{tab:S8}

\smallskip

\begin{tabular}{|l|c|c|c|} \hline
    &   Data & $S_8$  &  reference\\
$[1]$ & Planck TTTEEE  & $0.828\pm 0.013$ &  \cite{Efstathiou:2021} \\
$[2]$ & Planck TTTEEE+Planck lensing & $0.829\pm 0.012$ &  \cite{Efstathiou:2021} \\
$[3]$ & ACT lensing+BAO &   $0.840\pm 0.02$8 &  \cite{Madhavacheril:2024} \\
$[4]$ & ACT lensing$\times$unWISE ($z= 0.2-1.6$) &   $0.813\pm 0.021$ &  \cite{Farren:2024} \\ 
$[5]$ & ACT lensing+Planck lensing+unWISE 3$\times2$pt & $0.816 \pm 0.015$  &  \cite{Farren:2024b} \\ 
$[6]$ & Planck lensing$\times$DESI (LRG)  ($z - 0.4- 1.0$) & $0.762\pm 0.024$   &  \cite{Sailor:2024} \\        
$[7]$ & ACT lensing$\times$DESI (LRG)  ($z - 0.4- 1.0$)  & $0.790^{+0.024}_{-0.027}\ \ $   &  \cite{Sailor:2024} \\ 
$[8]$ &  DESI Full Shape ($ z = 0.2-2.1$)  &  $0.836 \pm 0.035$  & \cite{DESIFS:2024} \\ \hline
\end{tabular}
\end{center}
\end{table*}

  \medskip
  
  \noindent
  (A) A low value of $S_8$

  \smallskip
  
  As noted in Sects.~\ref{sec:Introduction} and \ref{sec:motivation}, the amplitude of the tSZ spectrum is
  sensitive to value of the $S_8$ parameter quantifying the amplitude of the mass fluctuation spectrum.
  Motivated by indications of a low value of $S_8$ from cosmic shear surveys, M23 ran a set of simulations
  (labelled LS8) of a \LCDM\ cosmology.  but with the amplitude of the fluctuation spectrum
  lowered to give $S_8 = 0.766$ (corresponding to the low $S_8$ `cosmic shear' cosmology discussed by \cite{Amon:2023}). 
  The tSZ spectrum of the \flamingo\ LS8 cosmology is plotted in Fig.~\ref{fig:multinest}. At $\ell = 2780$,
  the LS8 model predicts $10^{12}D^{yy}_{2870} = 0.90$, whereas the measured value from Table \ref{tab:yyspec}
is   $0.57 \pm 0.062$ (which may be an overestimate since we have neglected the kSZ effect). For comparison, the \flamingo\ \Planck\ \LCDM\ prediction is
  $10^{12}D^{yy}_{2870} = 1.69$. The scaling between these two predictions is in good
  agreement with  Eq.~\ref{equ:tSZscaling2}.  To match the ACT/SPT tSZ amplitude would require
  a value of $S_8 \sim 0.73$, which is lower than inferred from cosmic shear surveys  \cite[e.g.][]{DES+KIDS:2023}.

    Furthermore,  a number of new measurements sensitive to linear scales have been reported which  disfavour a low $S_8$  cosmology,  as summarized in Table \ref{tab:S8}. The \Planck\ lensing and ACT DR6  measurements
    give values of $S_8$ that are in excellent agreement with the values inferred from the \Planck\
    temperature and polarization measurements (entries [1]-[3]). The CMB lensing measurements are sensitive to
    the mass distribution over a broad range of redshifts peaked at $z \sim 2$. The redshift range can be sharpened by
    cross-correlating CMB lensing with galaxy surveys. Entries [4]-[7] report results cross-correlating ACT and Planck
    lensing measurements with the unWISE catalogue of infrared galaxies \citep{Schlafly:2019} and the DESI Luminous Red Galaxy (LRG) sample.  The final entry [8] summarizes the results of the full shape modeling of galaxy and quasar clustering from the first year  DESI observations,  which is sensitive to $S_8$ via redshift space distortions.  Measurements [4], [6]  and [8] are
    largely independent and if combined give $S_8 = 0.798 \pm 0.014$, which is within  $1.5 \sigma$ of the
    \Planck\ TTTEEE value in  entry [1]. It therefore seems extremely unlikely that a low value of $S_8$ is the reason
    that the \flamingo\ simulations fail to match the ACT/SPT tSZ measurements.

      \medskip
  
  \noindent
  (B) Enhanced Baryonic Feedback

  \smallskip

  Another possibility is that baryonic feedback is much more important than modeled in  the baseline \flamingo\ simulations.\footnote{The first version of this paper compared the empirical results shown in
  Fig.~\ref{fig:multinest} with halo models of the tSZ spectrum from \cite{Omori:2024} which attempted to model enhanced baryonic feedback. However, the \cite{Omori:2024}  models do not reproduce the tSZ spectra measured directly from the simulations 
  on which the models are based (see Fig. 6 of \cite{McCarthy:2018}).}
  The red line in Fig.~\ref{fig:multinest} shows results from a \flamingo\ simulation with strong baryonic feedback \citep{McCarthy:2024c}. For this model, the AGN feedback prescription was adjusted so that
 the gas fractions in groups are $8\sigma$ lower than in the baseline model (hence the designation $f_{\rm gas} -8\sigma$). Even in this case, the model fails to match the low tSZ power inferred from ACT and SPT.  The orange line in Fig.~\ref{fig:multinest} shows a model with the self-similar evolution parameter of Eq.~\ref{equ:PS1} set to $\epsilon=1$. This also fails to match the ACT/SPT measurements.

There is however  evidence to support the idea that the baseline \flamingo\ simulations are underestimating the effects of baryonic
feedback.  Planck+ACT measurements stacked on galaxy reconstructed velocities derived from the Baryon Oscillation Spectroscopic Survey   \citep{Schaan:2021} leads to a kSZ signal favouring higher levels of baryonic feedback than in the baseline \flamingo\ simulations \citep{Bigwood:2024, McCarthy:2024c}. Evidence for high levels of baryonic feedback has been presented by \citep{Hadzhiyska:2024} from a similar kSZ analysis using ACT maps stacked on DESI LRGs using photometric redshifts to infer the velocity field.
  We also note that cosmic shear tSZ cross-correlation measurements suggest that high levels of baryonic feedback
  are required to reconcile a \Planck\ \LCDM\ cosmology with observations \citep{Troster:2022, Pandey:2023, McCarthy:2023, laposta:2024}. 
However, despite  these results it remains an open question of whether baryonic feedback can explain  tSZ spectrum
at high multipoles deduced from ACT and SPT.

  Finally, we note that the tSZ power spectrum has been used in many papers to constrain cosmology,  largely neglecting the role of baryonic feedback \citep[e.g.][]{PCCS2:2016, Hurier:2017, Salvati:2018, Tanimura:2021, Tanimura:2023}. The results presented here suggest that  baryonic feedback is an essential ingredient in shaping the tSZ spectrum and cannot be ignored. 

  We note that two papers have appeared while this paper was in the final stages of revision: (i) The KiDS collaboration has published results on cosmic shear  from the KiDS-Legacy Survey \citep{KiDS_legacy:2025} which surveys $1347$ square degrees and extends the redshift reach to redshift $2$. Improvements in the photometric redshifts and various other aspects of the cosmic shear analysis lead to a shift in the $S_8$ measurement compared to earlier KiDS results with the new analysis finding $S_8 = 0.815^{+0.016}_{-0.021}$, consistent with the \Plancks \LCDM\ 
  value quoted in Table~\ref{tab:S8}. This results strengthens the conclusion that the observed tSZ spectrum cannot be explained by invoking a low value of $S_8$;  (ii) the ACT collaboration have published  power spectra from  ACT DR6 \citep{ACT_DR6:2025} at frequencies of $98$, $150$ and $220$ GHz. They solve for a tSZ contribution to the temperature power spectra as in earlier papers
  \citep{Choi:2020} using the tSZ template power spectrum $^{\rm Bat}_\ell$ from \cite{Battaglia:2012}. With the increased signal-to-noise of the ACT DR6 data the are able to solve for a shape parameter $\alpha_{\rm tSZ}$, such that $D^{yy} = a^{yy}_{\rm tSZ}(D^{\rm Bat}_\ell/D^{\rm Bat}_{3000}) (\ell/3000)^
  {\alpha_{\rm tSZ}}$. They find $a^{yy}_{\rm tSZ} = 0.49 \pm 0.06$ and $\alpha_{\rm tSZ}=-0.6 \pm 0.2$. Their best fit is plotted
  as the violet dashed line in Fig.~\ref{fig:multinest} and is consistent with the results presented in this paper. It would be interesting to perform a reconstruction of the tSZ spectrum using ACT-DR6.  The agrrement of our results with ACT-DR6 emphasises the need for further research to establish whether baronic feedback can lead to a tSZ power spectrum
  with the amplitude and shape shown in Fig.~\ref{fig:multinest} .

\section{Acknowledgements}
GPE is grateful to the  Leverhulme Foundation for the award of a Leverhulme Emeritus Fellowship.
FMcC acknowledges support from the European Research Council (ERC) under the European Union's Horizon 2020 research and innovation programme (Grant agreement No.~851274).  The Flatiron Institute is a division of the Simons Foundation.  We thank Alex Amon,  Leah Bigwood, Boris Bolliet,  Will Coulton,  Colin Hill, Hiranya Peiris, Calvin Preston, Joop Schaye and Debora Sijacki, for many useful discussions concerning this work. We thank the referee for their comments on the manuscript.

\section*{Data Availability} 

No new data were generated or analysed in support of this research.

\bibliographystyle{mnras}
\bibliography{tSZpaper} 

\begin{thebibliography}{}
\makeatletter
\relax
\def\mn@urlcharsother{\let\do\@makeother \do\$\do\&\do\#\do\^\do\_\do\%\do\~}
\def\mn@doi{\begingroup\mn@urlcharsother \@ifnextchar [ {\mn@doi@}
  {\mn@doi@[]}}
\def\mn@doi@[#1]#2{\def\@tempa{#1}\ifx\@tempa\@empty \href
  {http://dx.doi.org/#2} {doi:#2}\else \href {http://dx.doi.org/#2} {#1}\fi
  \endgroup}
\def\mn@eprint#1#2{\mn@eprint@#1:#2::\@nil}
\def\mn@eprint@arXiv#1{\href {http://arxiv.org/abs/#1} {{\tt arXiv:#1}}}
\def\mn@eprint@dblp#1{\href {http://dblp.uni-trier.de/rec/bibtex/#1.xml}
  {dblp:#1}}
\def\mn@eprint@#1:#2:#3:#4\@nil{\def\@tempa {#1}\def\@tempb {#2}\def\@tempc
  {#3}\ifx \@tempc \@empty \let \@tempc \@tempb \let \@tempb \@tempa \fi \ifx
  \@tempb \@empty \def\@tempb {arXiv}\fi \@ifundefined
  {mn@eprint@\@tempb}{\@tempb:\@tempc}{\expandafter \expandafter \csname
  mn@eprint@\@tempb\endcsname \expandafter{\@tempc}}}

\bibitem[\protect\citeauthoryear{{Addison}, {Dunkley}  \& {Spergel}}{{Addison}
  et~al.}{2012}]{Addison:2012}
{Addison} G.~E.,  {Dunkley} J.,   {Spergel} D.~N.,  2012, \mn@doi [\mnras]
  {10.1111/j.1365-2966.2012.21664.x}, \href
  {https://ui.adsabs.harvard.edu/abs/2012MNRAS.427.1741A} {427, 1741}

\bibitem[\protect\citeauthoryear{{Ade} et~al.,}{{Ade} et~al.}{2019}]{SO_2019}
{Ade} P.,  et~al., 2019, \mn@doi [\jcap] {10.1088/1475-7516/2019/02/056}, \href
  {https://ui.adsabs.harvard.edu/abs/2019JCAP...02..056A} {2019, 056}

\bibitem[\protect\citeauthoryear{{Amon} \& {Efstathiou}}{{Amon} \&
  {Efstathiou}}{2022}]{AmonEfstathiou:2022}
{Amon} A.,  {Efstathiou} G.,  2022, \mn@doi [\mnras] {10.1093/mnras/stac2429},
  \href {https://ui.adsabs.harvard.edu/abs/2022MNRAS.516.5355A} {516, 5355}

\bibitem[\protect\citeauthoryear{{Amon} et~al.,}{{Amon}
  et~al.}{2022}]{Amon:2022}
{Amon} A.,  et~al., 2022, \mn@doi [\prd] {10.1103/PhysRevD.105.023514}, \href
  {https://ui.adsabs.harvard.edu/abs/2022PhRvD.105b3514A} {105, 023514}

\bibitem[\protect\citeauthoryear{{Amon} et~al.,}{{Amon}
  et~al.}{2023}]{Amon:2023}
{Amon} A.,  et~al., 2023, \mn@doi [\mnras] {10.1093/mnras/stac2938}, \href
  {https://ui.adsabs.harvard.edu/abs/2023MNRAS.518..477A} {518, 477}

\bibitem[\protect\citeauthoryear{{Arnaud}, {Pratt}, {Piffaretti},
  {B{\"o}hringer}, {Croston}  \& {Pointecouteau}}{{Arnaud}
  et~al.}{2010}]{Arnaud:2010}
{Arnaud} M.,  {Pratt} G.~W.,  {Piffaretti} R.,  {B{\"o}hringer} H.,  {Croston}
  J.~H.,   {Pointecouteau} E.,  2010, \mn@doi [\aap]
  {10.1051/0004-6361/200913416}, \href
  {https://ui.adsabs.harvard.edu/abs/2010A&A...517A..92A} {517, A92}

\bibitem[\protect\citeauthoryear{{Asgari} et~al.,}{{Asgari}
  et~al.}{2021}]{Asgari:2021}
{Asgari} M.,  et~al., 2021, \mn@doi [\aap] {10.1051/0004-6361/202039070}, \href
  {https://ui.adsabs.harvard.edu/abs/2021A&A...645A.104A} {645, A104}

\bibitem[\protect\citeauthoryear{{Battaglia}, {Bond}, {Pfrommer}  \&
  {Sievers}}{{Battaglia} et~al.}{2012}]{Battaglia:2012}
{Battaglia} N.,  {Bond} J.~R.,  {Pfrommer} C.,   {Sievers} J.~L.,  2012,
  \mn@doi [\apj] {10.1088/0004-637X/758/2/75}, \href
  {https://ui.adsabs.harvard.edu/abs/2012ApJ...758...75B} {758, 75}

\bibitem[\protect\citeauthoryear{{B{\'e}thermin} et~al.,}{{B{\'e}thermin}
  et~al.}{2012}]{Bethermin:2012}
{B{\'e}thermin} M.,  et~al., 2012, \mn@doi [\apjl]
  {10.1088/2041-8205/757/2/L23}, \href
  {https://ui.adsabs.harvard.edu/abs/2012ApJ...757L..23B} {757, L23}

\bibitem[\protect\citeauthoryear{{Bigwood} et~al.,}{{Bigwood}
  et~al.}{2024}]{Bigwood:2024}
{Bigwood} L.,  et~al., 2024, \mn@doi [\mnras] {10.1093/mnras/stae2100}, \href
  {https://ui.adsabs.harvard.edu/abs/2024MNRAS.534..655B} {534, 655}

\bibitem[\protect\citeauthoryear{{Bleem} et~al.,}{{Bleem}
  et~al.}{2022}]{Bleem:2022}
{Bleem} L.~E.,  et~al., 2022, \mn@doi [\apjs] {10.3847/1538-4365/ac35e9}, \href
  {https://ui.adsabs.harvard.edu/abs/2022ApJS..258...36B} {258, 36}

\bibitem[\protect\citeauthoryear{{Bolliet}, {Comis}, {Komatsu}  \&
  {Mac{\'\i}as-P{\'e}rez}}{{Bolliet} et~al.}{2018}]{Bolliet:2018}
{Bolliet} B.,  {Comis} B.,  {Komatsu} E.,   {Mac{\'\i}as-P{\'e}rez} J.~F.,
  2018, \mn@doi [\mnras] {10.1093/mnras/sty823}, \href
  {https://ui.adsabs.harvard.edu/abs/2018MNRAS.477.4957B} {477, 4957}

\bibitem[\protect\citeauthoryear{{Braspenning} et~al.,}{{Braspenning}
  et~al.}{2023}]{Brasspenning:2023}
{Braspenning} J.,  et~al., 2023, \mn@doi [arXiv e-prints]
  {10.48550/arXiv.2312.08277}, \href
  {https://ui.adsabs.harvard.edu/abs/2023arXiv231208277B} {p. arXiv:2312.08277}

\bibitem[\protect\citeauthoryear{{Carlstrom}, {Holder}  \& {Reese}}{{Carlstrom}
  et~al.}{2002}]{Carlstrom:2002}
{Carlstrom} J.~E.,  {Holder} G.~P.,   {Reese} E.~D.,  2002, \mn@doi [\araa]
  {10.1146/annurev.astro.40.060401.093803}, \href
  {https://ui.adsabs.harvard.edu/abs/2002ARA&A..40..643C} {40, 643}

\bibitem[\protect\citeauthoryear{{Chandran}, {Remazeilles}  \&
  {Barreiro}}{{Chandran} et~al.}{2023}]{Chandran:2023}
{Chandran} J.,  {Remazeilles} M.,   {Barreiro} R.~B.,  2023, \mn@doi [\mnras]
  {10.1093/mnras/stad3156}, \href
  {https://ui.adsabs.harvard.edu/abs/2023MNRAS.526.5682C} {526, 5682}

\bibitem[\protect\citeauthoryear{Chen \& Wright}{Chen \&
  Wright}{2009}]{Chen:2008gw}
Chen X.,  Wright E.~L.,  2009, \mn@doi [Astrophys. J.]
  {10.1088/0004-637X/694/1/222}, 694, 222

\bibitem[\protect\citeauthoryear{{Chluba}, {Hill}  \& {Abitbol}}{{Chluba}
  et~al.}{2017}]{Chluba:2017}
{Chluba} J.,  {Hill} J.~C.,   {Abitbol} M.~H.,  2017, \mn@doi [\mnras]
  {10.1093/mnras/stx1982}, \href
  {https://ui.adsabs.harvard.edu/abs/2017MNRAS.472.1195C} {472, 1195}

\bibitem[\protect\citeauthoryear{{Choi} et~al.,}{{Choi}
  et~al.}{2020}]{Choi:2020}
{Choi} S.~K.,  et~al., 2020, \mn@doi [\jcap] {10.1088/1475-7516/2020/12/045},
  \href {https://ui.adsabs.harvard.edu/abs/2020JCAP...12..045C} {2020, 045}

\bibitem[\protect\citeauthoryear{{Cole} \& {Kaiser}}{{Cole} \&
  {Kaiser}}{1988}]{Cole:1988}
{Cole} S.,  {Kaiser} N.,  1988, \mn@doi [\mnras] {10.1093/mnras/233.3.637},
  \href {https://ui.adsabs.harvard.edu/abs/1988MNRAS.233..637C} {233, 637}

\bibitem[\protect\citeauthoryear{{Coulton} et~al.,}{{Coulton}
  et~al.}{2024}]{Coulton:2024}
{Coulton} W.,  et~al., 2024, \mn@doi [\prd] {10.1103/PhysRevD.109.063530},
  \href {https://ui.adsabs.harvard.edu/abs/2024PhRvD.109f3530C} {109, 063530}

\bibitem[\protect\citeauthoryear{{DESI Collaboration} et~al.,}{{DESI
  Collaboration} et~al.}{2024}]{DESIFS:2024}
{DESI Collaboration} et~al., 2024, \mn@doi [arXiv e-prints]
  {10.48550/arXiv.2411.12022}, \href
  {https://ui.adsabs.harvard.edu/abs/2024arXiv241112022D} {p. arXiv:2411.12022}

\bibitem[\protect\citeauthoryear{{Dark Energy Survey and Kilo-Degree Survey
  Collaboration} et~al.,}{{Dark Energy Survey and Kilo-Degree Survey
  Collaboration} et~al.}{2023}]{DES+KIDS:2023}
{Dark Energy Survey and Kilo-Degree Survey Collaboration} et~al., 2023, \mn@doi
  [The Open Journal of Astrophysics] {10.21105/astro.2305.17173}, \href
  {https://ui.adsabs.harvard.edu/abs/2023OJAp....6E..36D} {6, 36}

\bibitem[\protect\citeauthoryear{{Das} et~al.,}{{Das} et~al.}{2014}]{Das:2014}
{Das} S.,  et~al., 2014, \mn@doi [\jcap] {10.1088/1475-7516/2014/04/014}, \href
  {https://ui.adsabs.harvard.edu/abs/2014JCAP...04..014D} {2014, 014}

\bibitem[\protect\citeauthoryear{{Delabrouille}, {Cardoso}, {Le Jeune},
  {Betoule}, {Fay}  \& {Guilloux}}{{Delabrouille}
  et~al.}{2009}]{Delabrouille:2009}
{Delabrouille} J.,  {Cardoso} J.~F.,  {Le Jeune} M.,  {Betoule} M.,  {Fay} G.,
   {Guilloux} F.,  2009, \mn@doi [\aap] {10.1051/0004-6361:200810514}, \href
  {https://ui.adsabs.harvard.edu/abs/2009A&A...493..835D} {493, 835}

\bibitem[\protect\citeauthoryear{{Dunkley} et~al.,}{{Dunkley}
  et~al.}{2013}]{Dunkley:2013}
{Dunkley} J.,  et~al., 2013, \mn@doi [\jcap] {10.1088/1475-7516/2013/07/025},
  \href {https://ui.adsabs.harvard.edu/abs/2013JCAP...07..025D} {2013, 025}

\bibitem[\protect\citeauthoryear{{Efstathiou} \& {Gratton}}{{Efstathiou} \&
  {Gratton}}{2021}]{Efstathiou:2021}
{Efstathiou} G.,  {Gratton} S.,  2021, \mn@doi [The Open Journal of
  Astrophysics] {10.21105/astro.1910.00483}, \href
  {https://ui.adsabs.harvard.edu/abs/2021OJAp....4E...8E} {4, 8}

\bibitem[\protect\citeauthoryear{{Efstathiou} \& {Migliaccio}}{{Efstathiou} \&
  {Migliaccio}}{2012}]{Efstathiou:2012}
{Efstathiou} G.,  {Migliaccio} M.,  2012, \mn@doi [\mnras]
  {10.1111/j.1365-2966.2012.21059.x}, \href
  {https://ui.adsabs.harvard.edu/abs/2012MNRAS.423.2492E} {423, 2492}

\bibitem[\protect\citeauthoryear{{Everett} et~al.,}{{Everett}
  et~al.}{2020}]{Everett:2020}
{Everett} W.~B.,  et~al., 2020, \mn@doi [\apj] {10.3847/1538-4357/ab9df7},
  \href {https://ui.adsabs.harvard.edu/abs/2020ApJ...900...55E} {900, 55}

\bibitem[\protect\citeauthoryear{{Farren} et~al.,}{{Farren}
  et~al.}{2024a}]{Farren:2024b}
{Farren} G.~S.,  et~al., 2024a, \mn@doi [arXiv e-prints]
  {10.48550/arXiv.2409.02109}, \href
  {https://ui.adsabs.harvard.edu/abs/2024arXiv240902109F} {p. arXiv:2409.02109}

\bibitem[\protect\citeauthoryear{{Farren} et~al.,}{{Farren}
  et~al.}{2024b}]{Farren:2024}
{Farren} G.~S.,  et~al., 2024b, \mn@doi [\apj] {10.3847/1538-4357/ad31a5},
  \href {https://ui.adsabs.harvard.edu/abs/2024ApJ...966..157F} {966, 157}

\bibitem[\protect\citeauthoryear{{Feroz}, {Hobson}  \& {Bridges}}{{Feroz}
  et~al.}{2009}]{Feroz:2009}
{Feroz} F.,  {Hobson} M.~P.,   {Bridges} M.,  2009, \mn@doi [\mnras]
  {10.1111/j.1365-2966.2009.14548.x}, \href
  {https://ui.adsabs.harvard.edu/abs/2009MNRAS.398.1601F} {398, 1601}

\bibitem[\protect\citeauthoryear{{Feroz}, {Hobson}  \& {Bridges}}{{Feroz}
  et~al.}{2011}]{Feroz:2011}
{Feroz} F.,  {Hobson} M.~P.,   {Bridges} M.,  2011, {MultiNest: Efficient and
  Robust Bayesian Inference} (\mn@eprint {ascl} {1109.006})

\bibitem[\protect\citeauthoryear{{Hadzhiyska} et~al.,}{{Hadzhiyska}
  et~al.}{2024}]{Hadzhiyska:2024}
{Hadzhiyska} B.,  et~al., 2024, \mn@doi [arXiv e-prints]
  {10.48550/arXiv.2407.07152}, \href
  {https://ui.adsabs.harvard.edu/abs/2024arXiv240707152H} {p. arXiv:2407.07152}

\bibitem[\protect\citeauthoryear{{Heymans} et~al.,}{{Heymans}
  et~al.}{2012}]{Heymans:2012}
{Heymans} C.,  et~al., 2012, \mn@doi [\mnras]
  {10.1111/j.1365-2966.2012.21952.x}, \href
  {https://ui.adsabs.harvard.edu/abs/2012MNRAS.427..146H} {427, 146}

\bibitem[\protect\citeauthoryear{{Hikage} et~al.,}{{Hikage}
  et~al.}{2019}]{Hikage:2019}
{Hikage} C.,  et~al., 2019, \mn@doi [\pasj] {10.1093/pasj/psz010}, \href
  {https://ui.adsabs.harvard.edu/abs/2019PASJ...71...43H} {71, 43}

\bibitem[\protect\citeauthoryear{{Hill} \& {Pajer}}{{Hill} \&
  {Pajer}}{2013}]{2013PhRvD..88f3526H}
{Hill} J.~C.,  {Pajer} E.,  2013, \mn@doi [\prd] {10.1103/PhysRevD.88.063526},
  \href {https://ui.adsabs.harvard.edu/abs/2013PhRvD..88f3526H} {88, 063526}

\bibitem[\protect\citeauthoryear{{Hill} \& {Spergel}}{{Hill} \&
  {Spergel}}{2014}]{Hill:2014}
{Hill} J.~C.,  {Spergel} D.~N.,  2014, \mn@doi [\jcap]
  {10.1088/1475-7516/2014/02/030}, \href
  {https://ui.adsabs.harvard.edu/abs/2014JCAP...02..030H} {2014, 030}

\bibitem[\protect\citeauthoryear{{Hurier} \& {Lacasa}}{{Hurier} \&
  {Lacasa}}{2017}]{Hurier:2017}
{Hurier} G.,  {Lacasa} F.,  2017, \mn@doi [\aap] {10.1051/0004-6361/201630041},
  \href {https://ui.adsabs.harvard.edu/abs/2017A&A...604A..71H} {604, A71}

\bibitem[\protect\citeauthoryear{{Hurier}, {Mac{\'\i}as-P{\'e}rez}  \&
  {Hildebrandt}}{{Hurier} et~al.}{2013}]{Hurier:2013}
{Hurier} G.,  {Mac{\'\i}as-P{\'e}rez} J.~F.,   {Hildebrandt} S.,  2013, \mn@doi
  [\aap] {10.1051/0004-6361/201321891}, \href
  {https://ui.adsabs.harvard.edu/abs/2013A&A...558A.118H} {558, A118}

\bibitem[\protect\citeauthoryear{{Jenkins}, {Frenk}, {White}, {Colberg},
  {Cole}, {Evrard}, {Couchman}  \& {Yoshida}}{{Jenkins}
  et~al.}{2001}]{Jenkins:2001}
{Jenkins} A.,  {Frenk} C.~S.,  {White} S.~D.~M.,  {Colberg} J.~M.,  {Cole} S.,
  {Evrard} A.~E.,  {Couchman} H.~M.~P.,   {Yoshida} N.,  2001, \mn@doi [\mnras]
  {10.1046/j.1365-8711.2001.04029.x}, \href
  {https://ui.adsabs.harvard.edu/abs/2001MNRAS.321..372J} {321, 372}

\bibitem[\protect\citeauthoryear{{Komatsu} \& {Kitayama}}{{Komatsu} \&
  {Kitayama}}{1999}]{Komatsu:1999}
{Komatsu} E.,  {Kitayama} T.,  1999, \mn@doi [\apjl] {10.1086/312364}, \href
  {https://ui.adsabs.harvard.edu/abs/1999ApJ...526L...1K} {526, L1}

\bibitem[\protect\citeauthoryear{{Komatsu} \& {Seljak}}{{Komatsu} \&
  {Seljak}}{2002}]{Komatsu:2002}
{Komatsu} E.,  {Seljak} U.,  2002, \mn@doi [\mnras]
  {10.1046/j.1365-8711.2002.05889.x}, \href
  {https://ui.adsabs.harvard.edu/abs/2002MNRAS.336.1256K} {336, 1256}

\bibitem[\protect\citeauthoryear{Li, Puglisi, Madhavacheril  \& Alvarez}{Li
  et~al.}{2022}]{Li:2021ial}
Li Z.,  Puglisi G.,  Madhavacheril M.~S.,   Alvarez M.~A.,  2022, \mn@doi
  [JCAP] {10.1088/1475-7516/2022/08/029}, 08, 029

\bibitem[\protect\citeauthoryear{{Louis} et~al.,}{{Louis}
  et~al.}{2025}]{ACT_DR6:2025}
{Louis} T.,  et~al., 2025, \mn@doi [arXiv e-prints]
  {10.48550/arXiv.2503.14452}, \href
  {https://ui.adsabs.harvard.edu/abs/2025arXiv250314452L} {p. arXiv:2503.14452}

\bibitem[\protect\citeauthoryear{{MacCrann}, {Zuntz}, {Bridle}, {Jain}  \&
  {Becker}}{{MacCrann} et~al.}{2015}]{MacCrann:2015}
{MacCrann} N.,  {Zuntz} J.,  {Bridle} S.,  {Jain} B.,   {Becker} M.~R.,  2015,
  \mn@doi [\mnras] {10.1093/mnras/stv1154}, \href
  {https://ui.adsabs.harvard.edu/abs/2015MNRAS.451.2877M} {451, 2877}

\bibitem[\protect\citeauthoryear{{Madhavacheril} et~al.,}{{Madhavacheril}
  et~al.}{2020}]{Madhavacheril:2020}
{Madhavacheril} M.~S.,  et~al., 2020, \mn@doi [\prd]
  {10.1103/PhysRevD.102.023534}, \href
  {https://ui.adsabs.harvard.edu/abs/2020PhRvD.102b3534M} {102, 023534}

\bibitem[\protect\citeauthoryear{{Madhavacheril} et~al.,}{{Madhavacheril}
  et~al.}{2024}]{Madhavacheril:2024}
{Madhavacheril} M.~S.,  et~al., 2024, \mn@doi [\apj]
  {10.3847/1538-4357/acff5f}, \href
  {https://ui.adsabs.harvard.edu/abs/2024ApJ...962..113M} {962, 113}

\bibitem[\protect\citeauthoryear{{Mak}, {Challinor}, {Efstathiou}  \&
  {Lagache}}{{Mak} et~al.}{2017}]{Mak:2017}
{Mak} D. S.~Y.,  {Challinor} A.,  {Efstathiou} G.,   {Lagache} G.,  2017,
  \mn@doi [\mnras] {10.1093/mnras/stw3112}, \href
  {https://ui.adsabs.harvard.edu/abs/2017MNRAS.466..286M} {466, 286}

\bibitem[\protect\citeauthoryear{{Maniyar}, {B{\'e}thermin}  \&
  {Lagache}}{{Maniyar} et~al.}{2021}]{Maniyar:2021}
{Maniyar} A.,  {B{\'e}thermin} M.,   {Lagache} G.,  2021, \mn@doi [\aap]
  {10.1051/0004-6361/202038790}, \href
  {https://ui.adsabs.harvard.edu/abs/2021A&A...645A..40M} {645, A40}

\bibitem[\protect\citeauthoryear{{McCarthy} \& {Hill}}{{McCarthy} \&
  {Hill}}{2024}]{McCarthy:2024}
{McCarthy} F.,  {Hill} J.~C.,  2024, \mn@doi [\prd]
  {10.1103/PhysRevD.109.023528}, \href
  {https://ui.adsabs.harvard.edu/abs/2024PhRvD.109b3528M} {109, 023528}

\bibitem[\protect\citeauthoryear{{McCarthy}, {Le Brun}, {Schaye}  \&
  {Holder}}{{McCarthy} et~al.}{2014}]{McCarthy:2014}
{McCarthy} I.~G.,  {Le Brun} A.~M.~C.,  {Schaye} J.,   {Holder} G.~P.,  2014,
  \mn@doi [\mnras] {10.1093/mnras/stu543}, \href
  {https://ui.adsabs.harvard.edu/abs/2014MNRAS.440.3645M} {440, 3645}

\bibitem[\protect\citeauthoryear{{McCarthy}, {Bird}, {Schaye},
  {Harnois-Deraps}, {Font}  \& {van Waerbeke}}{{McCarthy}
  et~al.}{2018}]{McCarthy:2018}
{McCarthy} I.~G.,  {Bird} S.,  {Schaye} J.,  {Harnois-Deraps} J.,  {Font}
  A.~S.,   {van Waerbeke} L.,  2018, \mn@doi [\mnras] {10.1093/mnras/sty377},
  \href {https://ui.adsabs.harvard.edu/abs/2018MNRAS.476.2999M} {476, 2999}

\bibitem[\protect\citeauthoryear{{McCarthy} et~al.,}{{McCarthy}
  et~al.}{2023}]{McCarthy:2023}
{McCarthy} I.~G.,  et~al., 2023, \mn@doi [\mnras] {10.1093/mnras/stad3107},
  \href {https://ui.adsabs.harvard.edu/abs/2023MNRAS.526.5494M} {526, 5494}

\bibitem[\protect\citeauthoryear{{McCarthy} et~al.,}{{McCarthy}
  et~al.}{2024}]{McCarthy:2024c}
{McCarthy} I.~G.,  et~al., 2024, \mn@doi [arXiv e-prints]
  {10.48550/arXiv.2410.19905}, \href
  {https://ui.adsabs.harvard.edu/abs/2024arXiv241019905M} {p. arXiv:2410.19905}

\bibitem[\protect\citeauthoryear{{Omori}}{{Omori}}{2024}]{Omori:2024}
{Omori} Y.,  2024, \mn@doi [\mnras] {10.1093/mnras/stae1031}, \href
  {https://ui.adsabs.harvard.edu/abs/2024MNRAS.530.5030O} {530, 5030}

\bibitem[\protect\citeauthoryear{{Pandey} et~al.,}{{Pandey}
  et~al.}{2023}]{Pandey:2023}
{Pandey} S.,  et~al., 2023, \mn@doi [\mnras] {10.1093/mnras/stad2268}, \href
  {https://ui.adsabs.harvard.edu/abs/2023MNRAS.525.1779P} {525, 1779}

\bibitem[\protect\citeauthoryear{{Planck Collaboration} et~al.,}{{Planck
  Collaboration} et~al.}{2013}]{Planck_PS:2013}
{Planck Collaboration} et~al., 2013, \mn@doi [\aap]
  {10.1051/0004-6361/201220053}, \href
  {https://ui.adsabs.harvard.edu/abs/2013A&A...550A.133P} {550, A133}

\bibitem[\protect\citeauthoryear{{Planck Collaboration} et~al.,}{{Planck
  Collaboration} et~al.}{2014a}]{Planck_SR:2014}
{Planck Collaboration} et~al., 2014a, \mn@doi [\aap]
  {10.1051/0004-6361/201321531}, \href
  {https://ui.adsabs.harvard.edu/abs/2014A&A...571A...9P} {571, A9}

\bibitem[\protect\citeauthoryear{{Planck Collaboration} et~al.,}{{Planck
  Collaboration} et~al.}{2014b}]{Planck_params:2014}
{Planck Collaboration} et~al., 2014b, \mn@doi [\aap]
  {10.1051/0004-6361/201321591}, \href
  {https://ui.adsabs.harvard.edu/abs/2014A&A...571A..16P} {571, A16}

\bibitem[\protect\citeauthoryear{{Planck Collaboration} et~al.,}{{Planck
  Collaboration} et~al.}{2016a}]{Planck_Ymap:2016}
{Planck Collaboration} et~al., 2016a, \mn@doi [\aap]
  {10.1051/0004-6361/201525826}, \href
  {https://ui.adsabs.harvard.edu/abs/2016A&A...594A..22P} {594, A22}

\bibitem[\protect\citeauthoryear{{Planck Collaboration} et~al.,}{{Planck
  Collaboration} et~al.}{2016b}]{PCCS2:2016}
{Planck Collaboration} et~al., 2016b, \mn@doi [\aap]
  {10.1051/0004-6361/201526914}, \href
  {https://ui.adsabs.harvard.edu/abs/2016A&A...594A..26P} {594, A26}

\bibitem[\protect\citeauthoryear{{Planck Collaboration} et~al.,}{{Planck
  Collaboration} et~al.}{2020a}]{Planck_Likelihood:2020}
{Planck Collaboration} et~al., 2020a, \mn@doi [\aap]
  {10.1051/0004-6361/201936386}, \href
  {https://ui.adsabs.harvard.edu/abs/2020A&A...641A...5P} {641, A5}

\bibitem[\protect\citeauthoryear{{Planck Collaboration} et~al.,}{{Planck
  Collaboration} et~al.}{2020b}]{Planck_params:2020}
{Planck Collaboration} et~al., 2020b, \mn@doi [\aap]
  {10.1051/0004-6361/201833910}, \href
  {https://ui.adsabs.harvard.edu/abs/2020A&A...641A...6P} {641, A6}

\bibitem[\protect\citeauthoryear{{Planck Collaboration} et~al.,}{{Planck
  Collaboration} et~al.}{2020c}]{PR4:2020}
{Planck Collaboration} et~al., 2020c, \mn@doi [\aap]
  {10.1051/0004-6361/202038073}, \href
  {https://ui.adsabs.harvard.edu/abs/2020A&A...643A..42P} {643, A42}

\bibitem[\protect\citeauthoryear{Posta, Alonso, Chisari, Ferreira  \&
  García-García}{Posta et~al.}{2024}]{laposta:2024}
Posta A.~L.,  Alonso D.,  Chisari N.~E.,  Ferreira T.,   García-García C.,
  2024, $X+y$: insights on gas thermodynamics from the combination of X-ray and
  thermal Sunyaev-Zel'dovich data cross-correlated with cosmic shear
  (\mn@eprint {arXiv} {2412.12081}), \url {https://arxiv.org/abs/2412.12081}

\bibitem[\protect\citeauthoryear{{Preston}, {Amon}  \& {Efstathiou}}{{Preston}
  et~al.}{2023}]{Preston:2023}
{Preston} C.,  {Amon} A.,   {Efstathiou} G.,  2023, \mn@doi [\mnras]
  {10.1093/mnras/stad2573}, \href
  {https://ui.adsabs.harvard.edu/abs/2023MNRAS.525.5554P} {525, 5554}

\bibitem[\protect\citeauthoryear{{Reichardt} et~al.,}{{Reichardt}
  et~al.}{2012}]{Reichardt:2012}
{Reichardt} C.~L.,  et~al., 2012, \mn@doi [\apj] {10.1088/0004-637X/755/1/70},
  \href {https://ui.adsabs.harvard.edu/abs/2012ApJ...755...70R} {755, 70}

\bibitem[\protect\citeauthoryear{{Reichardt} et~al.,}{{Reichardt}
  et~al.}{2021}]{Reichardt:2021}
{Reichardt} C.~L.,  et~al., 2021, \mn@doi [\apj] {10.3847/1538-4357/abd407},
  \href {https://ui.adsabs.harvard.edu/abs/2021ApJ...908..199R} {908, 199}

\bibitem[\protect\citeauthoryear{Remazeilles, Delabrouille  \&
  Cardoso}{Remazeilles et~al.}{2011}]{Remazeilles:2010hq}
Remazeilles M.,  Delabrouille J.,   Cardoso J.-F.,  2011, \mn@doi [Mon. Not.
  Roy. Astron. Soc.] {10.1111/j.1365-2966.2010.17624.x}, 410, 2481

\bibitem[\protect\citeauthoryear{{Rosenberg}, {Gratton}  \&
  {Efstathiou}}{{Rosenberg} et~al.}{2022}]{Rosenberg:2022}
{Rosenberg} E.,  {Gratton} S.,   {Efstathiou} G.,  2022, \mn@doi [\mnras]
  {10.1093/mnras/stac2744}, \href
  {https://ui.adsabs.harvard.edu/abs/2022MNRAS.517.4620R} {517, 4620}

\bibitem[\protect\citeauthoryear{{Sailer} et~al.,}{{Sailer}
  et~al.}{2024}]{Sailor:2024}
{Sailer} N.,  et~al., 2024, \mn@doi [arXiv e-prints]
  {10.48550/arXiv.2407.04607}, \href
  {https://ui.adsabs.harvard.edu/abs/2024arXiv240704607S} {p. arXiv:2407.04607}

\bibitem[\protect\citeauthoryear{{Salvati}, {Douspis}  \& {Aghanim}}{{Salvati}
  et~al.}{2018}]{Salvati:2018}
{Salvati} L.,  {Douspis} M.,   {Aghanim} N.,  2018, \mn@doi [\aap]
  {10.1051/0004-6361/201731990}, \href
  {https://ui.adsabs.harvard.edu/abs/2018A&A...614A..13S} {614, A13}

\bibitem[\protect\citeauthoryear{{Schaan} et~al.,}{{Schaan}
  et~al.}{2021}]{Schaan:2021}
{Schaan} E.,  et~al., 2021, \mn@doi [\prd] {10.1103/PhysRevD.103.063513}, \href
  {https://ui.adsabs.harvard.edu/abs/2021PhRvD.103f3513S} {103, 063513}

\bibitem[\protect\citeauthoryear{{Schaye} et~al.,}{{Schaye}
  et~al.}{2023}]{Schaye:2023}
{Schaye} J.,  et~al., 2023, \mn@doi [\mnras] {10.1093/mnras/stad2419}, \href
  {https://ui.adsabs.harvard.edu/abs/2023MNRAS.526.4978S} {526, 4978}

\bibitem[\protect\citeauthoryear{{Schlafly}, {Meisner}  \& {Green}}{{Schlafly}
  et~al.}{2019}]{Schlafly:2019}
{Schlafly} E.~F.,  {Meisner} A.~M.,   {Green} G.~M.,  2019, \mn@doi [\apjs]
  {10.3847/1538-4365/aafbea}, \href
  {https://ui.adsabs.harvard.edu/abs/2019ApJS..240...30S} {240, 30}

\bibitem[\protect\citeauthoryear{{Secco} et~al.,}{{Secco}
  et~al.}{2022}]{Secco:2022}
{Secco} L.~F.,  et~al., 2022, \mn@doi [\prd] {10.1103/PhysRevD.105.023515},
  \href {https://ui.adsabs.harvard.edu/abs/2022PhRvD.105b3515S} {105, 023515}

\bibitem[\protect\citeauthoryear{{Shaw}, {Zahn}, {Holder}  \&
  {Dor{\'e}}}{{Shaw} et~al.}{2009}]{2009ApJ...702..368S}
{Shaw} L.~D.,  {Zahn} O.,  {Holder} G.~P.,   {Dor{\'e}} O.,  2009, \mn@doi
  [\apj] {10.1088/0004-637X/702/1/368}, \href
  {https://ui.adsabs.harvard.edu/abs/2009ApJ...702..368S} {702, 368}

\bibitem[\protect\citeauthoryear{{Sievers} et~al.,}{{Sievers}
  et~al.}{2013}]{Sievers:2013}
{Sievers} J.~L.,  et~al., 2013, \mn@doi [\jcap]
  {10.1088/1475-7516/2013/10/060}, \href
  {https://ui.adsabs.harvard.edu/abs/2013JCAP...10..060S} {2013, 060}

\bibitem[\protect\citeauthoryear{Stein, Alvarez, Bond, van Engelen  \&
  Battaglia}{Stein et~al.}{2020}]{Stein:2020its}
Stein G.,  Alvarez M.~A.,  Bond J.~R.,  van Engelen A.,   Battaglia N.,  2020,
  \mn@doi [JCAP] {10.1088/1475-7516/2020/10/012}, 10, 012

\bibitem[\protect\citeauthoryear{{Sunyaev} \& {Zeldovich}}{{Sunyaev} \&
  {Zeldovich}}{1972}]{SZ:1972}
{Sunyaev} R.~A.,  {Zeldovich} Y.~B.,  1972, Comments on Astrophysics and Space
  Physics, \href {https://ui.adsabs.harvard.edu/abs/1972CoASP...4..173S} {4,
  173}

\bibitem[\protect\citeauthoryear{{Tanimura}, {Douspis}, {Aghanim}  \&
  {Salvati}}{{Tanimura} et~al.}{2022}]{Tanimura:2021}
{Tanimura} H.,  {Douspis} M.,  {Aghanim} N.,   {Salvati} L.,  2022, \mn@doi
  [\mnras] {10.1093/mnras/stab2956}, \href
  {https://ui.adsabs.harvard.edu/abs/2022MNRAS.509..300T} {509, 300}

\bibitem[\protect\citeauthoryear{{Tanimura}, {Douspis}  \&
  {Aghanim}}{{Tanimura} et~al.}{2023}]{Tanimura:2023}
{Tanimura} H.,  {Douspis} M.,   {Aghanim} N.,  2023, in {Ruffino} R.,
  {Vereshchagin} G.,  eds, The Sixteenth Marcel Grossmann Meeting. On Recent
  Developments in Theoretical and Experimental General Relativity,
  Astrophysics, and Relativistic Field Theories. pp 1527--1531,
  \mn@doi{10.1142/9789811269776_0121}

\bibitem[\protect\citeauthoryear{Thorne, Dunkley, Alonso  \& Naess}{Thorne
  et~al.}{2017}]{Thorne:2016ifb}
Thorne B.,  Dunkley J.,  Alonso D.,   Naess S.,  2017, \mn@doi [Mon. Not. Roy.
  Astron. Soc.] {10.1093/mnras/stx949}, 469, 2821

\bibitem[\protect\citeauthoryear{{Tr{\"o}ster} et~al.,}{{Tr{\"o}ster}
  et~al.}{2022}]{Troster:2022}
{Tr{\"o}ster} T.,  et~al., 2022, \mn@doi [\aap] {10.1051/0004-6361/202142197},
  \href {https://ui.adsabs.harvard.edu/abs/2022A&A...660A..27T} {660, A27}

\bibitem[\protect\citeauthoryear{{Tucci}, {Toffolatti}, {de Zotti}  \&
  {Mart{\'\i}nez-Gonz{\'a}lez}}{{Tucci} et~al.}{2011}]{Tucci:2011}
{Tucci} M.,  {Toffolatti} L.,  {de Zotti} G.,   {Mart{\'\i}nez-Gonz{\'a}lez}
  E.,  2011, \mn@doi [\aap] {10.1051/0004-6361/201116972}, \href
  {https://ui.adsabs.harvard.edu/abs/2011A&A...533A..57T} {533, A57}

\bibitem[\protect\citeauthoryear{{Viero} et~al.,}{{Viero}
  et~al.}{2013}]{Viero:2013}
{Viero} M.~P.,  et~al., 2013, \mn@doi [\apj] {10.1088/0004-637X/772/1/77},
  \href {https://ui.adsabs.harvard.edu/abs/2013ApJ...772...77V} {772, 77}

\bibitem[\protect\citeauthoryear{{Wright} et~al.,}{{Wright}
  et~al.}{2025}]{KiDS_legacy:2025}
{Wright} A.~H.,  et~al., 2025, \mn@doi [arXiv e-prints]
  {10.48550/arXiv.2503.19441}, \href
  {https://ui.adsabs.harvard.edu/abs/2025arXiv250319441W} {p. arXiv:2503.19441}

\makeatother
\end{thebibliography}
\end{document}